\documentclass[preprint,superscriptaddress,nofootinbib]{revtex4}
\usepackage[dvips]{graphicx}

\usepackage[usenames]{color}
\usepackage{setspace}

\usepackage{amsmath}
\usepackage{amssymb}
\usepackage{amsthm}

\newcommand{\VEV}[1]{\left\langle #1 \right\rangle}

\newcommand{\nn}{\nonumber}
\newcommand{\diag}{{\rm diag}}

\newcommand{\TeV}{\mbox{TeV}}
\newcommand{\GeV}{\mbox{GeV}}

\newcommand{\ie}{{\it i.e.}}

\newcommand{\Z}[1]{{\mathbb Z}_#1}
\newcommand{\abs}[1]{\left| #1 \right|}
\newcommand{\s}[1]{\tilde{#1}}

\newcommand{\tr}{{\rm {tr}}}
\newcommand{\supP}[1]{^{(#1)}}

\newcommand{\bequ}{\begin{equation}}
\newcommand{\eequ}{\end{equation}}
\newcommand{\beqn}{\begin{eqnarray}}
\newcommand{\eeqn}{\end{eqnarray}}
\newcommand{\bctr}{\begin{center}}
\newcommand{\ectr}{\end{center}}
\newcommand{\bit}{\begin{itemize}}
\newcommand{\eit}{\end{itemize}}
\newcommand{\Ls}{\left(}
\newcommand{\Rs}{\right)}

\def\lromn#1{\uppercase\expandafter{\romannumeral#1}}

\def\GeV{\rm{GeV}}
\def\TeV{\rm{TeV}}

\begin{document}

  \preprint{UT-HET-085}


  \title{Higgs as a Probe of Supersymmetric Grand Unification with the
    Hosotani Mechanism}
  \author{Mitsuru Kakizaki}
  \email{kakizaki@sci.u-toyama.ac.jp}
  \affiliation{
  Department of Physics,
  University of Toyama, Toyama 930-8555, Japan
  }
  \author{Shinya Kanemura}
  \email{kanemu@sci.u-toyama.ac.jp}
  \affiliation{
  Department of Physics,
  University of Toyama, Toyama 930-8555, Japan
  }
  \author{Hiroyuki Taniguchi} 
  \email{taniguchi@jodo.sci.u-toyama.ac.jp}
  \affiliation{
  Department of Physics,
  University of Toyama, Toyama 930-8555, Japan
  }
  \author{Toshifumi Yamashita}
  \email{tyamashi@aichi-med-u.ac.jp}
  \affiliation{
    Department of Physics,
    Aichi Medical University, Nagakute 480-1195, Japan
  }
  \begin{abstract}
    The supersymmetric grand unified theory where the $SU(5)$ gauge
    symmetry is broken by the Hosotani mechanism predicts the
    existence of adjoint chiral superfields whose masses are at the
    supersymmetry breaking scale.  The Higgs sector is extended with
    the $SU(2)_L$ triplet with hypercharge zero and neutral singlet
    chiral multiplets from that in the minimal supersymmetric standard
    model.  Since the triplet and singlet chiral multiplets originate
    from a higher-dimensional vector multiplet, this model is highly
    predictive.  Properties of the particles in the Higgs sector are
    characteristic and can be different from those in the Standard
    Model and other models.  We evaluate deviations in coupling
    constants of the standard model-like Higgs boson and the mass
    spectrum of the additional Higgs bosons.  We find that our model
    is discriminative from the others by precision measurements of
    these coupling constants and masses of the additional Higgs
    bosons.  This model can be a good example of grand unification
    that is testable at future collider experiments such as the
    luminosity up-graded Large Hadron Collider and future
    electron-positron colliders.
  \end{abstract}
\maketitle


\cleardoublepage
\pagenumbering{arabic}

\section{Introduction}

One of the most prominent achievements in particle physics in the past
decades is discovery of a new boson whose mass is around 125 GeV, as
reported in 2012 by the ATLAS and CMS collaborations of the CERN Large
Hadron Collider (LHC) \cite{LHC}.  After that, properties of the new
particle have been carefully investigated, and turned out to be
consistent with those of the Standard Model (SM) Higgs boson.  Now the
SM has been established as a successful low energy effective theory
that can consistently describe phenomena below the energy scale of
${\cal O}(100)$ GeV.

However, several high energy experiments and cosmological observations
show evidences for new physics beyond the SM, which include neutrino
oscillations, existence of dark matter and baryon asymmetry of the
universe.  In addition to such experimental results, the SM suffers
from theoretical problems.  One is a serious fine-tuning problem
called the hierarchy problem.  To reproduce the weak scale Higgs boson
mass, huge cancellation between its bare mass and contribution from
radiative corrections is required.  Another is that the reason why the
electric charges of the SM particles are fractionally quantized is
unexplained.

It is intriguing that some of theoretical problems can be elegantly
solved by introducing concepts of supersymmetry (SUSY) and grand
unification \cite{GUT,SUSY-GUT}.  The SUSY offers us a solution to
the hierarchy problem.  The quadratically divergent contributions to
the Higgs boson mass from the SM particles are canceled if we
introduce their partner particles whose spins differ from those of the
corresponding SM particles by half.  Grand Unified Theories (GUTs)
provide unified descriptions of the SM gauge groups.  Simultaneously,
SM fermion multiples are embedded into larger group representations,
leading to the charge quantization.  Therefore, combination of the
SUSY and the grand unification is an excellent candidate for the
underlying theory.  Moreover, in the minimal SUSY GUT, the three gauge
coupling constants are naturally unified at a high energy scale.

Although the idea itself is fascinating, GUT models have several
difficulties.  Notice that the typical energy scale of the gauge coupling
unification (GCU) in conventional SUSY GUTs is around $10^{16}$ GeV.
Given such a high GUT scale, superheavy GUT particles completely decouple
from the low energy effective theory \cite{Appelquist:1974tg}.
Therefore, testing GUTs usually relies on checking relations among
masses and coupling constants at the TeV scale, which are related to
each other through renormalization group equations (RGEs).  Moreover,
there is a fine tuning problem about the mass splitting between the
electroweak Higgs doublets and colored Higgs triplets, and many ideas
to solve the doublet-triplet (DT) splitting have been proposed
\cite{DW,SlidingSinglet,missingPARTN,pNG,orbifoldGUTs,Kakizaki:2001en}.  In
extended SUSY GUT models, the successful GCU is spoiled in many cases
and the GCU becomes a constraint instead of a prediction.

Recently, a SUSY GUT model that circumvents the above mentioned
difficulties is proposed by one of the authors \cite{gGHU-DTS} by
supersymmetrizing the Grand Gauge-Higgs Unification (GHU) \cite{gGHU},
where the GCU is just a constraint as in many extended SUSY GUT
models.  We call the supersymmetric version the Supersymmetric Grand
Gauge-Higgs Unification (SGGHU) in this paper.  The idea of the Grand
GHU is to break the GUT gauge group by applying the so-called Hosotani
mechanism \cite{hosotani}.  In the SGGHU, by using non-trivial vacuum
expectation value (VEV) of a Wilson loop, the doublet-triplet
splitting problem is naturally solved.  As a by-product, existence of
new light chiral adjoints is predicted.  At the TeV scale, our model
is reduced to the Minimal Supersymmetric Standard Model (MSSM) with a
color octet superfield, an $SU(2)_L$ triplet superfield with
hypercharge zero and a neutral singlet superfield.  In particular,
since the Higgs sector is extended by the triplet and singlet
superfields, we can test our GUT model by exploring properties of the
extended Higgs sector with collider experiments.  Due to couplings
between the MSSM Higgs doublets and the new Higgs triplet and singlet,
the SM-like Higgs boson mass can be more naturally as large as 125
GeV, as compared to the prediction of the MSSM
\cite{MSSMHiggsMass,MSSMHiggsMass2}.  Thus, the little hierarchy can
also be relaxed.  As we see later, even when the masses of the triplet
and singlet superfields are as small as the electroweak scale, it
turns out that the mass of the color octet is too large to probe its
effects at colliders due to radiative effects.

In this paper, we focus on the Higgs sector of the SGGHU, and explore
its phenomenological consequences.  We derive values of parameters in
the low energy effective theory using the RGEs, and evaluate how the
masses and couplings of the SGGHU Higgs sector particles are modified
from those in the MSSM due to the existence of the light triplet and
singlet chiral multiplets.  We emphasize that by measuring the masses
and couplings of the Higgs bosons precisely at the LHC and future
electron-positron colliders such as the International Linear Collider (ILC) 
\cite{ILC} and the CLIC \cite{CLIC}, particle physics models can be
distinguished.  We show that the SGGHU is a good example to show
the capabilities of collider experiments for testing GUT scale
physics.

This paper is organized as follows.  In Sec.~\ref{Sec:Model}, we
briefly review the model of the SGGHU and its low energy effective
theory.  Particular attention is paid to the Higgs sector, which is
extended by the triplet and singlet chiral multiplets.
Sec.~\ref{Sec:RGEAnalysis} is devoted to 
the discussion of the SM-like Higgs boson mass
using RGEs.  Some benchmark points reproducing the observed Higgs 
boson mass are provided.  In Sec.~\ref{Sec:Higgs},
predictions about couplings of the SM-like Higgs boson and mass
relation of additional Higgs bosons are presented based on the
benchmark points.  Definitions of model parameters and RGEs are
collected in Appendix~\ref{Sec:RGE}.  Mass matrices of Higgs bosons,
neutralinos and charginos are summarized in
Appendix~\ref{Sec:Higgsmass}.  Necessary formulae for computing
radiative corrections to the SM-like Higgs boson mass are also given
there.

\section{Model}
\label{Sec:Model}

\subsection{Review of Supersymmetric Grand Gauge-Higgs Unification}

In this subsection, we briefly review the grand GHU scenario proposed
in Ref.~\cite{gGHU-DTS}.  This scenario is a kind of the grand
unification where the Hosotani mechanism~\cite{hosotani} is employed
to break the $SU(5)$ unified gauge symmetry.  The simplest setup that
can accommodate the chiral fermions is a five-dimensional (5D) $SU(5)$
model compactified on an $S^1/\Z2$ orbifold with its radius being of
the GUT scale.  We first discuss the non-SUSY version of the simplest
setup discussed in Ref.~\cite{gGHU} for illustration purpose, and then
supersymmetrize it~\cite{gGHU-DTS}.

The Hosotani mechanism is a mechanism for gauge symmetry breaking
which works on higher-dimensional gauge theories.  To be more
concrete, the zero modes of extra-dimensional components of the gauge
fields, which behave as scalar fields after the compactification,
develop VEVs to break the gauge symmetry.  In order to apply this
mechanism to the $SU(5)$ unified gauge symmetry breaking, massless
adjoint scalar fields, with respect to the $SU(5)$ symmetry that
remains unbroken against the boundary conditions (BCs), should appear.
It is known that such components tend to be projected out in models
that realize the chiral fermions due to the orbifold BCs.
In Ref.~\cite{gGHU}, this difficulty is evaded via the
so-called diagonal embedding method~\cite{DiagonalEmbedding} which is
proposed in the context of the string theory.
In our field theoretical setup on the $S^1/\Z2$ orbifold, we impose
two copies of the gauge symmetry with an additional discrete symmetry
that exchanges the two gauge symmetries.  Namely, the symmetry is
$SU(5)\times SU(5)\times \Z2$ in our $SU(5)$ model.  Here, we name the
gauge fields for the two $SU(5)$ groups $A\supP1_M$ and $A\supP2_M$,
respectively, where $M=\mu(=0\mbox{-}3),5$ is a 5D Lorentzian index,
and define the eigenstates of the $\Z2$ action as
$X^{(\pm)}=(X\supP1\pm X\supP2)/\sqrt2$.  We set the BCs around the
two endpoints of the $S^1/\Z2$, $y_0=0$ and $y_\pi=\pi R$, as
\bequ
A_\mu\supP1(y_i-y) = A_\mu\supP2(y_i+y),\qquad A_5\supP1(y_i-y) =
-A_5\supP2(y_i+y),
\label{BCinHosotani}
\eequ
for $i=0,\pi$, where $y$ denotes the 5th dimensional coordinate.  With
these BCs, we see that $A_\mu\supP+$ and $A_5\supP-$ obey the Neumann
BC at each endpoint to have the zero-modes, and thus that the gauge
symmetry remaining unbroken in the 4D effective theory is the diagonal
part of the $SU(5)\times SU(5)$ (or our GUT symmetry is {\it embedded}
into the {\it diagonal} part) and an adjoint scalar field is actually
realized.

An interesting point is that the $A_5\supP-$ is not a simple adjoint
scalar field but composes a Wilson loop since it is a part of the
gauge field.  The Wilson loop is given by 
\bequ 
W={\cal P}\exp\Ls
i\int_{0}^{2\pi R} \frac{g}{\sqrt2}{{A_5\supP-}^a(T_1^a-T_2^a)} dy\Rs
\to \exp\Ls
i\diag\Ls\theta_1,\theta_2,\theta_3,\theta_4,\theta_5\Rs\Rs, 
\eequ
where ${\cal P}$ denotes the path-ordered integral, $g$ is the common
gauge coupling constant, $T_1$ and $T_2$ are the generators of the two
$SU(5)$ symmetries, and $a$ is an $SU(5)$ adjoint index.  In the last
expression, we show the expression on the fundamental representation
for concreteness, and we have used the (remaining) $SU(5)$ rotation to
diagonalize $A_5\supP-$.  This expression shows that the VEV (and
actually the system itself) is invariant under the shift
$\theta_i\to\theta_i+2\pi$.

The form of the VEV which is discussed in Ref.~\cite{gGHU} and which
we are interested in is given by $\theta_1=\theta_2=\theta_3=2\pi$ and
$\theta_4=\theta_5=-3\pi$, \ie\ $\VEV{W}=\diag(1,1,1,-1,-1)\equiv
P_W$.  This VEV does not affect the triplet component of the ${\bf5}$
representation but does affect the doublet to {\it split} them.  This
``missing VEV", which is forbidden for a simple adjoint scalar field
by the traceless condition, is allowed since the Wilson loop is valued
on a group instead of an algebra and thus is free from the condition.
This fact plays an essential role to solve the DT splitting problem.

In this paper, for simplicity, we do not consider matter fields that
are non-singlet under the both gauge groups.  We introduce for
instance a fermion $\Psi({\bf R},{\bf1})\supP1$ with ${\bf R}$ being a
representation of the $SU(5)$ group and its $\Z2$ partner $\Psi({\bf
  1},{\bf R})\supP2$.  Here, we call the above pair a "bulk ${\bf R}$
multiplet".  Their BCs are given as $ \Psi\supP1(y_\pi-y) =
-\eta_i^{\Psi}\gamma_5\Psi\supP2(y_\pi+y) $ where $\eta_i=\pm1$ is a
parameter associated with each fermion.  As one of $\eta_i$ can be
reabsorbed by changing $\gamma_5$, \ie\ by the charge conjugation, we
set $\eta_0=+1$ and $\eta_\pi=\eta$ hereafter.  Then, $\Psi_L\supP+$
and $\Psi_R\supP-$ have the zero-modes when $\eta=+1$ while they do
not when $\eta=-1$, when the VEV of $A_5$ vanishes.

Notice that it is always possible to gauge away the VEV of $A_5$
($\propto\theta$).  In this basis, called the Scherk-Schwartz basis,
the $SU(5)$ breaking effect appears only on the BCs as
\bequ
\Psi\supP1(y_\pi-y) = -\eta_i^{\Psi}\gamma_5 W_{\bf
  R}\Psi\supP2(y_\pi+y),
\label{FermionBCinSS}
\eequ 
where $W_{\bf R}$ is the Wilson line phase acting on ${\bf R}$.  In
concrete, for ${\bf R}={\bf 5}$ with $\eta=-1$ when the above VEV
$\VEV W=P_W$ is realized, the doublet component has the zero-mode
while the triplet does not.


The same story as discussed above can be applied also to the SUSY
extensions if we replace all the fields by the corresponding
superfields.  Thus, once the desired VEV $P_W$ is obtained, the DT
splitting is easily realized by introducing a bulk ${\bf 5}$
hypermultiplet with $\eta=-1$ for the MSSM Higgs fields.  In a similar
way, if we introduce bulk ${\bf 10}$ hypermultiplets with $\eta=+1$,
light vector-like pairs $(U,\bar U)$ ($(\bar {\bf3},{\bf1})_{-2/3}$)
and $(E,\bar E)$ ($({\bf1},{\bf1})_{1}$) appear, where the values
denote $SU(3)_C, SU(2)_L$ and $U(1)_Y$ quantum numbers.  This is
utilized to recover the gauge coupling unification later.

We note that the zero-modes appear always in vector-like pairs from
the bulk fields.  The chiral fermions can be put simply on each
boundary.  Interestingly, when the VEV $\VEV W=P_W$ is realized, bulk
fields serve vector-like pairs in $SU(5)$ incomplete multiplets while
the boundary fields which do not couple to $A_5$ and thus neither to
the $SU(5)$ breaking do chiral fermions in $SU(5)$ full multiplets.

The remaining task to show that the DT splitting problem is actually
solved is to examine when the VEV is realized.  Here, we do not
request that the vacuum resides on the global minimum but just require
only that it is stable so that the lifetime is long enough.  For this
purpose, we have to check if there is no huge tadpole term for the
fluctuation of $\theta_i$ around the desired vacuum, $\delta\theta_i$,
and if it is not tachyonic around the desired vacuum.  Since there are
two largely different scales, the compactification scale and the SUSY
breaking scale, the RG analysis should be performed.

Before going on the low energy effective theory, we note that the
exchanging $\Z2$ symmetry, under which $\delta\theta_i$ is odd,
remains unbroken on the relevant vacuum even though $\theta_i$ is
non-trivial.  This is understood by the transformation of the Wilson
line which is the order parameter.  Under the $\Z2$ action $W$
transforms as $W\to W^*$ and the VEV $\VEV{W}$ is invariant since it
is real.  This $\Z2$ invariance prohibit the tadpole terms.  In the
following, we introduce soft $\Z2$ breaking as small as the SUSY
breaking scale, and thus a small tadpole term will be generated.

\subsection{Low Energy Effective Theory}

As a consequence of the supersymmetrization of the grand gauge-Higgs
unification, there appear adjoint chiral superfields whose gauge
quantum numbers are the same as the SM gauge bosons.
Since these new adjoint fields are originally embedded in the
five-dimensional vector multiplets, their masses vanish in the SUSY
limit.  The chiral adjoints acquire masses after the SUSY breaking.
Therefore, typical masses of the adjoint supermultiplets 
are of the order of the SUSY breaking scale irrelevantly to
the compactification scale.

The low energy effective theory contains $SU(3)_C$ octet, $SU(2)_L$
triplet and singlet chiral superfields in addition to the MSSM
superfields.  As discussed later, since the mass of the octet chiral
superfield is $O(10)$ TeV due to the radiative correction, its effect
on the TeV scale phenomenology is negligible.  Therefore, we here
focus on the impact of the Higgs sector with the
$SU(2)_L$ triplet and singlet chiral superfields.

The Higgs sector is composed of the superfields shown in
Tab.~\ref{tb:fields}.  Here, $H_u$ ($H_d$) gives masses to the up-type
quarks (down-type quarks and charged leptons).  The superpotential of
the effective theory of our model is given by
\begin{eqnarray} \label{eq:WHiggs}
  W=\mu H_u \cdot H_d+\mu_\Delta^{}{\rm tr}(\Delta^2)+\frac{\mu_S^{}}2S^2
  +\lambda_\Delta^{} H_u \cdot \Delta H_d + \lambda_S^{}SH_u \cdot H_d \, ,
\end{eqnarray}
where $\Delta = \Delta^a \sigma^a/2$
with $\sigma^a (a=1,2,3)$ being the Pauli matrices.
Notice that there are no trilinear self-couplings among $S$ and
$\Delta$ although such couplings are not prohibited by the symmetry of
the effective theory because $S$ and $\Delta$ originate from the gauge
supermultiplet.  Moreover, the two new Higgs couplings
$\lambda_\Delta^{}$ and $\lambda_S^{}$ are unified with the unified gauge
coupling $g_{\rm GUT}$ as $\lambda_\Delta^{} =2\sqrt{5/3}\lambda_S^{} =
g_{\rm GUT}$ at the GUT scale.  Thus, this model is predictive
up to the soft SUSY breaking parameters.  Masses of the fermionic
components of $S$ and $\Delta$ are denoted by $\mu_S^{}$ and
$\mu_\Delta^{}$, respectively, and their magnitudes are of the order of
the TeV scale because they are generated due to the SUSY breaking
\cite{Burdman:2002se}.  Similarly, the supersymmetric tadpole parameter
of $S$ is expected to be of the order of $\mu m_{\rm SUSY}$, as
discussed in the previous section.  This tadpole term is removed by
field redefinition without loss of generality. 
The soft SUSY breaking terms are written by
\begin{eqnarray} \label{eq:VsoftHiggs}
  V_{\rm soft} 
  &=& \widetilde{m}_{H_d}^2 |H_d|^2 +\widetilde{m}_{H_u}^2|H_u|^2
  + 2 \widetilde{m}_\Delta^2 {\rm tr} (\Delta^\dag \Delta)
  +\widetilde{m}_S^2|S|^2 \nonumber \\
  &&+\left[B\mu H_u\cdot H_d +\xi S 
    + B_\Delta^{} \mu_\Delta^{}{\rm tr}(\Delta^2) +\frac{1}{2}B_S^{} \mu_S^{} S^2 \right. \nonumber \\
  &&\hspace{1.5em} \left. +\lambda_\Delta^{}A_\Delta^{} H_u\cdot \Delta H_d +\lambda_S^{} A_S^{} S H_u\cdot H_d +{\rm h.c.} \phantom{\frac{1}{2}}\right].
\end{eqnarray}
The low energy values of these parameters introduced in the Higgs
sector are obtained by solving the RGEs, which are discussed in the
next section.  It should be also noted that the VEV of the neutral
component of the triplet Higgs boson $v_\Delta^{}$ has to be smaller than
$\simeq 10~{\rm GeV}$ in order to satisfy the rho parameter
constraint.  

\begin{table}[t]
  \begin{tabular}{|c|c|c|c|}
    \hline
              & $SU(3)_C$ &  $SU(2)_L$ & $U(1)_Y$  \\ \hline \hline
    $H_u$     & ${\bf 1}$ &  ${\bf 2}$ & $+1/2$    \\ \hline
    $H_d$     & ${\bf 1}$ &  ${\bf 2}$ & $-1/2$    \\ \hline
    $\Delta$  & ${\bf 1}$ &  ${\bf 3}$ & $0$       \\ \hline
    $S$       & ${\bf 1}$ &  ${\bf 1}$ & $0$       \\ 
    \hline
  \end{tabular}
  \caption{\footnotesize 
    $SU(3)_C\times SU(2)_L\times U(1)_Y$ quantum numbers of 
    the Higgs sector superfields $H_u$, $H_d$, $\Delta$ and $S$.}
  \label{tb:fields}
\end{table}

\section{Reproduction of the Higgs Boson Mass}
\label{Sec:RGEAnalysis}

In this section, we discuss the mass of the SM-like Higgs boson based
on RG evolution of the coupling constants and the mass parameters in
our model.  First, we focus on the unification of the three gauge
coupling constants.  The existence of the light adjoint chiral
multiplets disturbs successful gauge coupling unification, which is
achieved in the minimal SUSY $SU(5)$ GUT.  In our model, extra
incomplete $SU(5)$ matter multiplets can be introduced so that the
gauge coupling unification is recovered \cite{gGHU-DTS}.  Next, we
derive values of the model parameters at the TeV scale by solving the
RGEs.  We show some benchmark points consistent with the observed
value of the mass of the Higgs boson.

\subsection{Coupling Unification}

\begin{figure}[t]
\centering
\includegraphics[width=120mm]{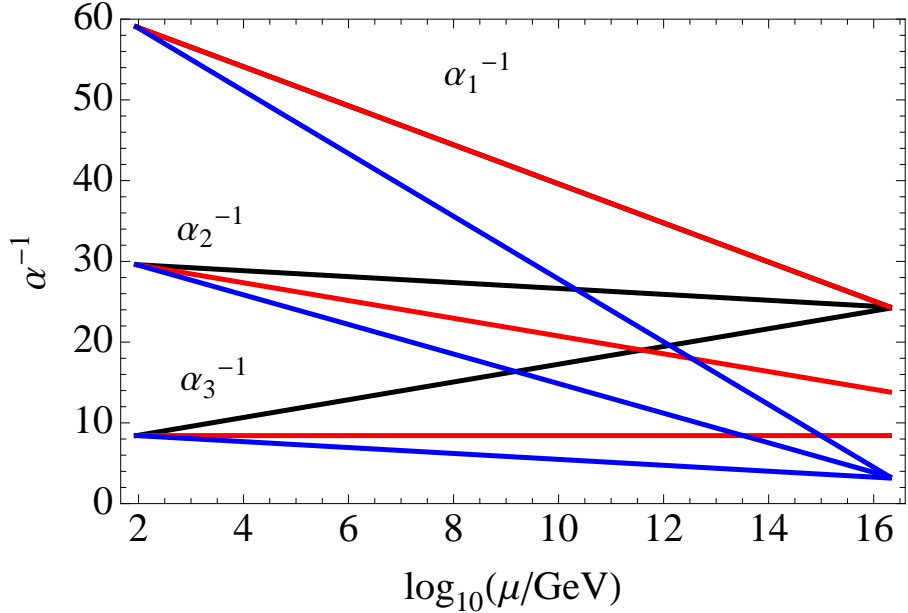}
\caption{\footnotesize Evolution of the gauge coupling constants in
  the MSSM (black lines), the MSSM with the adjoint multiplets (red),
  and the MSSM with the adjoint and additional chiral multiplets
  (blue).}
\label{fig:alphainv}
\end{figure}

The coefficients of the beta functions of the gauge couplings 
in the MSSM are given by
\begin{eqnarray}
  b_{\rm MSSM} = (33/5, 1, -3)\,  ,
\end{eqnarray}
while contributions from the adjoint chiral multiplets are
\begin{eqnarray}
 \delta_{\rm adj} b = (0, 2, 3)\, .
\end{eqnarray}
One way to recover the gauge coupling unification is to introduce
incomplete $SU(5)$ multiplets whose contributions are
\begin{eqnarray}
 \delta_{\rm add} b = (3+n, 1+n, n)\, ,
\end{eqnarray}
with $n$ being a natural number.  However, too large $n$ may cause
violation of perturbativity around the GUT scale. We here take $n=1$,
and the unified gauge coupling is in a perturbative region: $\alpha_G
\simeq 0.3$.  This case is realized by adding two vectorlike pairs of
$(\bar{L},L)$ $(({\bf1},{\bf2})_{-1/2})$, one of $(\bar{U},U)$ $((\bar
{\bf3},{\bf1})_{-2/3})$ and one of $(\bar{E},E)$
$(({\bf1},{\bf1})_{1})$ \cite{gGHU-DTS}.  Fig.~\ref{fig:alphainv}
shows evolution of the gauge coupling constants in the MSSM (black
lines), the MSSM with the adjoint multiplets (red), and the MSSM with
the adjoint and additional chiral multiplets (blue).
In this figure, we set the SUSY-breaking scale as the weak scale for 
simplicity.

In this model, the strong interaction is not asymptotically free
irrelevantly to the choice of the additional fields to recover the
gauge coupling unification.  Thus, the QCD corrections are large, and
the masses of the colored particles tend to be large at the TeV scale,
as compared to those in the MSSM.  It is interesting to examine the
extraordinary pattern of the mass spectrum of the colored particles
for the hadron colliders.  We, however, focus on the colorless fields;
the $SU(2)_L$ triplet and singlet Higgs multiplets.  These additional
fields couple to the two MSSM Higgs doublets.  Their coupling
constants push up the SM-like Higgs boson mass due to the tree level
$F$-term contribution, and thus the correct value of the Higgs boson
mass (around $125~{\rm GeV}$) can be easily realized.

Furthermore, they cause mixing between the MSSM doublet Higgs fields
and the additional Higgs fields, which results in modification of the
coupling constants of the SM-like Higgs field.  When such corrections
are large enough to be detected at collider experiments, we can
discriminate our model from other models by precisely measuring the
pattern of the deviations in the Higgs coupling constants.  In the
next section, we will discuss these issues in more details.

One of the characteristic features of this model is that the coupling
constants of the triplet and singlet Higgs multiplets are unified with
the SM gauge coupling constants at the GUT scale.  Thus, the
low-energy values of these coupling constants in the Higgs sector are
unambiguously determined by the RG running once the extra matters are
specified.

\begin{figure}[t]
\centering
\includegraphics[width=120mm]{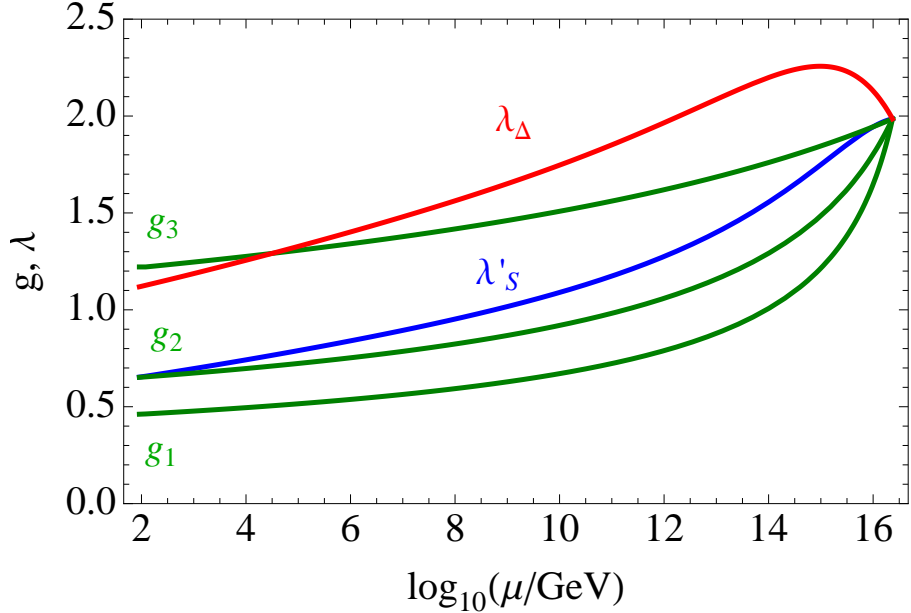}
\caption{\footnotesize An example of running of the Higgs triplet and
  singlet coupling constants $\lambda_\Delta^{}$ (red line) and
  $\lambda'_S$ (blue) as well as the gauge coupling constants $g_3$,
  $g_2$ and $g_1$ (green) from the top to the bottom.  The horizontal
  axis is the common logarithm of the energy scale in units of GeV.
  Here, we normalize the singlet and $U(1)_Y$ gauge couplings as
  $\lambda'_S=(2\sqrt{5/3}) \lambda_S^{}$ and $g_1=(\sqrt{5/3})g'$,
  respectively, and one loop RGEs are used. }
\label{fig:couplings}
\end{figure}

For instance, taking the above example of the additional chiral matter
multiplets to recover the gauge coupling unification, the Higgs sector
coupling constants $\lambda_\Delta^{}$ (red line) and $\lambda_S^{}$ (blue),
and the gauge coupling constants $g_{3,2,1}$ (green) evolve as shown
in Fig.~\ref{fig:couplings}.  Here, we normalize the singlet coupling
as $\lambda'_S=(2\sqrt{5/3}) \lambda_S^{}$, and the $U(1)_Y$ gauge
coupling as $g_1=(\sqrt{5/3})g'$, respectively, and one loop RGEs are
used.  For the list of the RGEs, see Appendix A.  Since the $SU(2)_L$
gauge coupling is strong around the GUT scale, $\lambda_\Delta^{}$ grows
as the energy decreases.  After the $SU(2)_L$ gauge coupling becomes
weak, $\lambda_\Delta^{}$ decreases as the energy decreases due to large
trilinear couplings in the superpotential.
We note that the triplet coupling $\lambda_\Delta^{}$ remains in a
perturbative region down to the TeV scale.  At the TeV scale, we
obtain
\begin{equation} \label{eq:couplings}
  \lambda_\Delta^{}=1.1\, , \qquad \lambda_S^{}=0.25\, .
\end{equation}
Similarly, the $\mu$-parameters of the adjoint chiral multiplets are
unified at the GUT scale, and their ratio at the TeV scale is
determined as $\mu_S^{}: \mu_\Delta^{}: \mu_O^{} = 1:2.9:230$, where $\mu_O^{}$
stands for the octet $\mu$-parameter.  The mass scale of the octet is
far beyond the reach of collider experiments, as discussed
qualitatively above.

Let us turn to the running of the soft SUSY breaking parameters.
Since the unified gauge coupling is strong, the gaugino masses around
the GUT scale must be large in order to avoid the experimental gluino
mass limit \cite{LHCgluino}.  For instance, for the unified gaugino
mass of $M_{1/2}= 3600$ GeV, the gluino mass is pushed down to
$m_{\tilde{g}}=1400$ GeV.  As a result, soft mass parameters at the
TeV scale are typically as large as $4$-$7$ TeV for colored particles
and $1$-$2$ TeV for colorless particles.  As in the MSSM, the soft
mass squared of the up-type Higgs boson has a large contribution due
to the large top Yukawa interaction.  Therefore, some tuning is needed
to realize electroweak symmetry breaking.  The higgsino mass parameter
$\mu$ and the CP-odd Higgs boson mass $m_A^{}$ also tend to be $1$-$4$
TeV.  In order to realize scenarios where some of the extra Higgs
boson masses are of the order of ${\cal O}(100)$ GeV, further tuning
is required among the input parameters.

\subsection{Benchmark Points and the Mass of the SM-like Higgs boson}

After the electroweak symmetry breaking, we obtain four CP-even, three
CP-odd and three charged Higgs bosons as physical states in the Higgs
sector, as well as six neutralinos and three charginos.  Features of
our model include new additional particles to the MSSM, and
differences in the properties of the MSSM Higgs bosons.
Among them, we here focus on the mass of the SM-like Higgs boson,
which is determined by low energy soft SUSY breaking parameters
obtained by solving the RGEs discussed above.

Before we discuss the cases where effects of the RG running is
involved in calculating the SM-like Higgs boson mass, we exemplify
rough predictions of our low energy effective theory without solving
the RG equations.  For relatively large triplet and singlet scalar masses,
the SM-like Higgs boson mass is approximately written as
\cite{Espinosa}
\begin{align} \label{eq:higgs_mass}
m_h^2 \simeq
m_Z^2 \cos^2 \beta
+\frac{3 m_t^4}{2\pi^2 v^2}\left( \ln{\frac{m_{\tilde{t}}^2}{m_t^2}} +
  \frac{X_t^2}{m_{\tilde{t}}^2} \left(1 -\frac{X_t^2}{12m_{\tilde{t}}^2} 
    \right) \right)
+\frac{1}{8}\lambda_\Delta^2 v^2 \sin^2 {2\beta}
+\frac{1}{2}\lambda_S^2 v^2 \sin^2 {2\beta}
\, ,
\end{align}
where $m_Z$ is the $Z$-boson mass, $m_t$ is the top quark mass,
$m_{\tilde{t}}$ is the average of the two stop masses, and $X_t = A_t
- \mu \cot \beta$ parametrizes mixing between the two stops.  The
first two terms correspond to the MSSM prediction.  The last two terms
originate from the existence of the trilinear couplings between the
MSSM Higgs doublets and the additional triplet and singlet.

Within the MSSM, at the tree level the SM-like Higgs boson mass is
smaller than the $Z$-boson mass.  In order to reach $125~\GeV$ using
the effect of the stop loop correction, the mass scale of the stops or
the mixing parameter $X_t$ should be very large.  For $X_t=0$, the
stop mass should be of the order of $O(10)~\TeV$.  Even in the maximum
mixing case where $X_t =\pm \sqrt{6}m_{\tilde{t}}$, the stop mass is
required to be as large as $O(1)~\TeV$\cite{MSSMHiggsMass2}.  We also
note that preferable range for $\tan \beta$ is larger than 10.

In our model, on the contrary, the predicted Higgs boson mass tends to
be larger than that in the MSSM thanks to the tree level $F$-term
contributions, in particular, for small $\tan \beta$ region.  Such a
result is reminiscent of the next-to-MSSM
(NMSSM\cite{NMSSM}), where the SM-like Higgs boson mass is
lifted up by coupling with a singlet superfield.

For computation of the masses of the Higgs scalars and superparticle,
we have used the public numerical code \texttt{SuSpect}
\cite{suspect}, which takes the $\overline{\rm DR}$ renormalization
scheme, instead of the approximate formula Eq.(\ref{eq:higgs_mass}).
We have appropriately modified \texttt{SuSpect} to add the new
contributions from the Higgs trilinear couplings.  Here, for
simplicity, we have taken the limit $v_\Delta^{} \to 0$.  The computation
of the SM-like Higgs boson mass including these triplet and singlet
contributions is described in Appendix B.  The LHC result
$m_h=125~{\rm GeV}$ can be achieved even for small $\tan \beta$ and
small stop mixing.  We note that the formula given in
Eq.(\ref{eq:higgs_mass}) is valid when the neutral components of the
triplet and singlet are heavier than the MSSM-like CP-even Higgs
bosons.  In general, the CP-even Higgs bosons mix with each other and
the formulae for their mass eigenvalues are rather complicated.

Next, let us consider the mass of the SM-like Higgs boson including the
radiative effects.  As we mentioned, in order to have a successful
electroweak symmetry breaking, fine tuning for input parameters at the
GUT scale is required.  Therefore, we will show some benchmark points
that reproduce the mass of the SM-like Higgs boson, instead of
scanning the parameter space.  We focus on the following three
different cases:
\begin{itemize}
\item[(A)] All the Higgs bosons other than the SM-like Higgs boson are heavy.
\item[(B)] The new Higgs bosons other than the MSSM-like Higgs bosons are heavy.
\item[(C)] The new Higgs bosons affect the SM-like Higgs boson couplings.
\end{itemize}
Bearing the fact that there are a few GeV uncertainties in the
numerical computation of the SM-like Higgs boson mass, we take the
range of $122~{\rm GeV} < m_h < 129~{\rm GeV}$ as its allowed region.
Examples of successful benchmark points of input parameters at the GUT
scale are listed in Tab. \ref{tab:benchmark-GUT}.  Here, $\mu$ and $B$
parameters for the extra matters have insignificant effects on Higgs
sector parameters, and are omitted from the list.  Values of
parameters of the TeV-scale effective theory are obtained after RG
running and shown in Tab. \ref{tab:benchmark-TeV}.  Definitions of the
parameters are provided in Appendix \ref{Sec:RGE}.

\begin{table}[t]
  \begin{flushleft}
{\footnotesize
  \begin{tabular}{|c||c|c|c|c|c|c|c|}
    \hline
    Case &
    $\tan \beta$ &
    $M_{1/2}$ &
    $\mu_{\Sigma}$
    \\ \hline  \hline
    (A)(B)(C) &
    $3$ &
    $3600~{\rm GeV}$ &
    $-300~{\rm GeV}$
    \\ \hline 
  \end{tabular}
  \begin{tabular}{|c||c|c|c|c|c|c|c|}
    \hline
    Case &
    $A_0$ & 
    $\widetilde{m}_0^2$ & 
    $\widetilde{m}_{H_u}^2$ &
    $\widetilde{m}_{H_d}^2$ &
    $\widetilde{m}_{5}^2$ &
    $\widetilde{m}_{10}^2$ &
    $\widetilde{m}_\Sigma^2$
    \\ \hline  \hline
    (A) &
    $5500~{\rm GeV}$ & 
    $(1000~{\rm GeV})^2$ &
    $(10375~{\rm GeV})^2$ &
    $(8570~{\rm GeV})^2$ & 
    $- (6300~{\rm GeV})^2$ &
    $- (2000~{\rm GeV})^2$ &
    $- (570~{\rm GeV})^2$
    \\ \hline 
    (B) &
    $1000~{\rm GeV}$ & 
    $(1800~{\rm GeV})^2$ &
    $(12604~{\rm GeV})^2$ &
    $(10381.5~{\rm GeV})^2$ & 
    $- (7700~{\rm GeV})^2$ &
    $- (1960~{\rm GeV})^2$ &
    $- (670~{\rm GeV})^2$
    \\ \hline 
    (C) &
    $8000~{\rm GeV}$ & 
    $(3000~{\rm GeV})^2$ &
    $(10605.1~{\rm GeV})^2$ &
    $(8751.4~{\rm GeV})^2$ & 
    $- (6418~{\rm GeV})^2$ &
    $- (1638.5~{\rm GeV})^2$ &
    $- (400~{\rm GeV})^2$
        \\ \hline 
  \end{tabular}
}
\end{flushleft}
  \caption{\footnotesize Benchmark points of input parameters at the 
    GUT scale.}
  \label{tab:benchmark-GUT}
\end{table}

\begin{table}[t]
  \begin{flushleft}
    {\footnotesize
  \begin{tabular}{|c||c|c|c|c|c|c|c|c|}
    \hline
    Case &
    $M_1$ &
    $M_2$ &
    $M_3$ &
    $\mu_{\Delta}$ &
    $\mu_S^{}$
    \\ \hline  \hline
    (A)(B)(C) &
    $194~{\rm GeV}$ & 
    $388~{\rm GeV}$ &
    $1360~{\rm GeV}$ &
    $-252~{\rm GeV}$ &
    $-85.8~{\rm GeV}$
    \\ \hline 
\end{tabular}
  \begin{tabular}{|c||c|c|c|c|c|c|c|c|}
    \hline
    Case &
    $\mu$ &
    $B \mu$ & 
    $\widetilde{m}_{u_3}$ & 
    $\widetilde{m}_{q_3}$ &
    $y_t A_t$
    \\ \hline  \hline
    (A) &
    $205~{\rm GeV}$ &
    $41400~{\rm GeV}^2$ &
    $3290~{\rm GeV}$ & 
    $4830~{\rm GeV}$ &
    $4030~{\rm GeV}$
    \\ \hline 
    (B) &
    $177~{\rm GeV}$ &
    $40800~{\rm GeV}^2$ &
    $1730~{\rm GeV}$ & 
    $4480~{\rm GeV}$ &
    $6050~{\rm GeV}$
    \\ \hline 
    (C) &
    $174~{\rm GeV}$ &
    $42000~{\rm GeV}^2$ &
    $4220~{\rm GeV}$ & 
    $5550~{\rm GeV}$ &
    $2910~{\rm GeV}$
    \\ \hline 
  \end{tabular}
  \begin{tabular}{|c||c|c|c|c|c|c||c|}
    \hline
    Case &
    $\widetilde{m}_{\Delta}$ & 
    $\widetilde{m}_{S}$ &
    $\lambda_\Delta^{} A_\Delta^{}$ &
    $\lambda'_S A_S^{}$ &
    $B_\Delta^{} \mu_\Delta^{}$ &
    $B_S^{} \mu_S^{}$ &
    $m_h$
    \\ \hline  \hline
    (A) &
    $607~{\rm GeV}$ & 
    $805~{\rm GeV}$ &
    $662~{\rm GeV}$ &
    $683~{\rm GeV}$ &
    $92000~{\rm GeV}^2$ &
    $-78700~{\rm GeV}^2$ &
    $123~{\rm GeV}$
    \\ \hline 
    (B) &
    $784~{\rm GeV}$ & 
    $612~{\rm GeV}$ &
    $1340~{\rm GeV}$ &
    $1110~{\rm GeV}$ &
    $30700~{\rm GeV}^2$ &
    $-110000~{\rm GeV}^2$ &
    $123~{\rm GeV}$
    \\ \hline 
    (C) &
    $521~{\rm GeV}$ & 
    $216~{\rm GeV}$ &
    $284~{\rm GeV}$ &
    $446~{\rm GeV}$ &
    $207000~{\rm GeV}^2$ &
    $-33600~{\rm GeV}^2$ &
    $122~{\rm GeV}$
    \\ \hline 
  \end{tabular}
}
  \end{flushleft}
  \caption{\footnotesize Parameters of the TeV-scale effective theory
    obtained after RG running.}
  \label{tab:benchmark-TeV}
\end{table}

\section{Impact on Higgs Properties}
\label{Sec:Higgs}

In this section, we discuss properties of the particles in the Higgs
sector.  We will show that our model can be distinguished from other
new physics models by measuring the masses and the coupling constants
of the Higgs sector particles at the LHC and future electron-positron
colliders \cite{HiggsWG,ILC,CLIC,ILCHiggs}.  Even in the cases where
the additional Higgs particles are beyond the reach of direct
discovery at these colliders, the existence of these new particles can
be indirectly probed by precise measurements of the coupling constants
of the discovered SM-like Higgs boson and MSSM Higgs boson masses.

\subsection{Vertices of the SM-like Higgs boson}

First, we address the couplings between the SM-like Higgs boson and SM
particles, which have been already measured to some extent at the LHC.
So far, no deviation that obviously contradicts the SM predictions
has been reported.  In the future, precision of these observables
will be significantly improved by the high-luminosity LHC and the ILC,
and therefore this method serves as a powerful tool in discriminating
beyond-the-SM models.

Discarding the VEV of the triplet Higgs boson, the Higgs boson
coupling with the $W$- or $Z$-boson is given by
\begin{align}
  g_{h V V}^{} = g_V^{} m_V^{} (R_{11}^S \cos \beta + R_{12}^S \sin \beta)\, ,
\quad V=W,Z
\end{align}
those with the up-type quarks, down-type quarks and charged leptons by
\begin{eqnarray} \label{eq:ghff}
g_{huu}^{} = \frac{\sqrt{2} m_u^{}}{v} \frac{R_{12}^S}{\sin \beta}\, , \quad
g_{hdd}^{} = \frac{\sqrt{2} m_d^{}}{v} \frac{R_{11}^S}{\cos \beta}\, , \quad
g_{h\ell\ell}^{} = \frac{\sqrt{2} m_\ell^{}}{v} \frac{R_{11}^S}{\cos \beta}\, ,
\end{eqnarray}
respectively, and the Higgs self-coupling is 
\begin{align}
g_{h h h}^{} = 
\sum_{a,b,c} R_{1a}^{S} R_{1b}^{S} R_{1c}^{S} \lambda_{s_a s_b s_c}^{}\, ,
\end{align}
where $R^S$ denotes the orthogonal matrix that diagonalizes the CP-even
Higgs mass matrix, and $\lambda_{s_a s_b s_c}$ are tree-level couplings 
among CP-even Higgs bosons in the gauge basis.
Their definitions are summarized in Appendix B.
The effective vertex of $h\gamma\gamma$ including contributions
from the additional charged Higgs bosons is given by
\begin{eqnarray}
  g_{h\gamma \gamma}^{} = 
  \sum_{f} N_c^{} Q_f^2 g_{hff}^{} A_{1/2}^{} (\tau_f^{}) 
  +  g_{hWW}^{} A_1^{} (\tau_W^{}) 
  + \sum_{h^\pm_i}
  \frac{m_W^2 \lambda_{hh^+_ih^-_i}^{}}{2 c_W^2 m_{h^\pm_i}^2}
  A_0^{} (\tau_{h^\pm_i}^{})\, ,
\end{eqnarray}
where the number of color is $N_c^{}=3$, and $Q_f^{}$ denote the electric
charges of fermions $f$.  For the definitions of the amplitudes $A_i^{}$,
see, for example, Ref.~\cite{Djouadi:2005gj}.
The Higgs boson couplings with the charged Higgs bosons are given by
\begin{eqnarray}
  \lambda_{h h^+_i h^-_i}^{} = 
  \sum_{a,b,c} R_{1a}^{S} U_{ib}^{C*} U_{ic}^{C} \lambda_{s_a w^+_b w^-_c}^{}\, .
\end{eqnarray}
The definitions of the unitary matrix $U^{C}$ and couplings 
$\lambda_{s_a w^+_b w^-_c}^{}$ are summarized in Appendix B.

The corresponding couplings in the SM are
\footnote{Since the Higgs trilinear coupling is calculated at the tree
  level, we choose $g_{hhh}^{}|_{\rm SM}^{} = m_Z^2/v$ for its
  normalization.}
\begin{eqnarray}
  && {g_{hVV}^{}|_{\rm SM}^{}} = g_V m_V, \quad
  {g_{huu}^{}|_{\rm SM}^{}} = \frac{\sqrt{2} m_u^{}}{v}\, ,\quad
  {g_{hdd}^{}|_{\rm SM}^{}} = \frac{\sqrt{2} m_d^{}}{v}\, ,
  \nonumber \\
  && {g_{h\ell\ell}^{}|_{\rm SM}^{}} = \frac{\sqrt{2} m_\ell^{}}{v}\, ,\quad
  {g_{hhh}^{}|_{\rm SM}^{}} = \frac{m_Z^2}{v}\, .
\end{eqnarray}
It is useful to define deviation parameters
\begin{eqnarray}
  \kappa_X^{} = \frac{g_{hXX}^{}}{g_{hXX}^{}|_{\rm SM}^{}} \, ,
\end{eqnarray}
where $X$ denotes SM particles.  Such deviations are extracted from
measurements of the decay widths of the Higgs boson.

\begin{figure}[t]
\includegraphics[width=100mm]{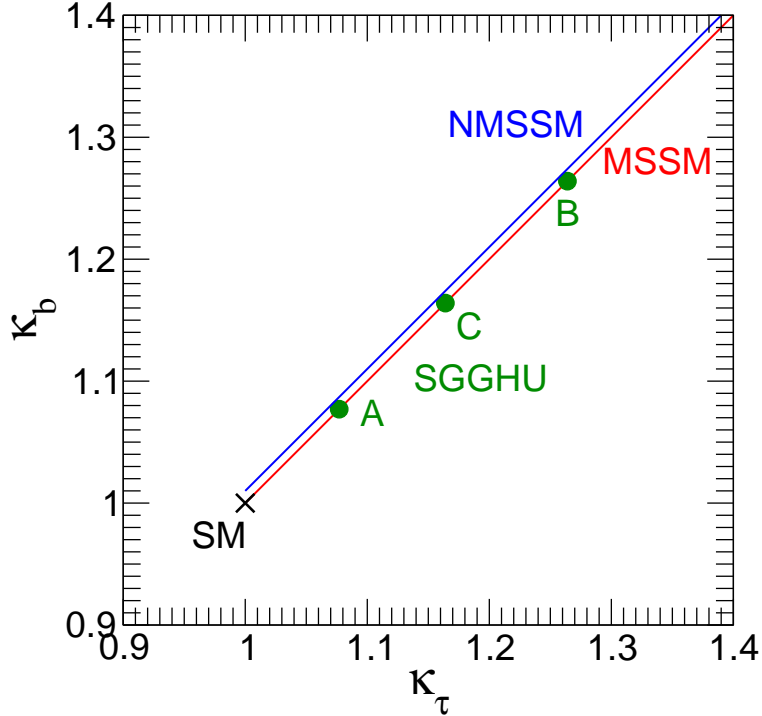}
\caption{\footnotesize The deviations in the Higgs boson coupling with
  the tau lepton $\kappa_\tau^{}$ and that with the bottom quark
  $\kappa_b^{}$ from the SM predictions are plotted.  The predictions
  of the three benchmark points (A), (B) and (C) in the SGGHU are
  shown with green blobs.  The MSSM and NMSSM predictions are shown
  with red and blue lines, respectively.  For the purpose of
  illustration, the NMSSM line is slightly displaced from
  $\kappa_\tau^{} = \kappa_b^{}$. }
\label{fig:taub}
\end{figure}

In Fig.~\ref{fig:taub}, the deviations in the Higgs boson coupling
with the tau lepton $\kappa_\tau^{}$ and that with the bottom quark
$\kappa_b^{}$ from the SM predictions are plotted.  The predictions of
the three benchmark points (A), (B) and (C) in the SGGHU are shown
with green blobs.  The MSSM and NMSSM predictions are shown with red
and blue lines, respectively.  Here, we simply adjust the stop masses
and mixing so that the observed Higgs boson mass is reproduced.  In
our model, the Higgs boson couplings to the down-type quarks and charged
leptons are common and fall in the category of the two Higgs doublet
model.  Therefore, the predicted SGGHU deviations lie on the MSSM and
NMSSM lines, as is evident from Eq.(\ref{eq:ghff}).  The recent LHC
results show strong evidence of the Higgs boson coupling with the tau lepton 
consistent with the SM prediction \cite{LHCtau}.  At the
ILC with $\sqrt{s}=500~{\rm GeV}$, expected accuracies for the
deviations $\kappa_\tau^{}$ and $\kappa_b^{}$ are 2.3\% and 1.6\%,
respectively \cite{ILCHiggs}.

\begin{figure}[t]
\includegraphics[width=100mm]{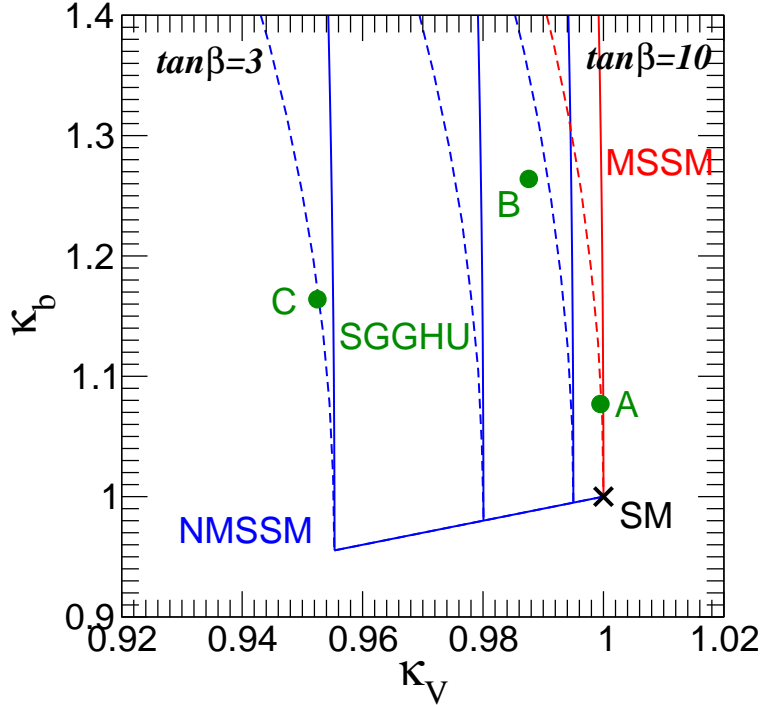}
\caption{\footnotesize The deviations in the Higgs boson coupling with
  the weak gauge bosons $\kappa_V^{}$ and that with the bottom quark
  $\kappa_b^{}$ from the SM predictions are plotted.  The predictions
  of the three benchmark points (A), (B) and (C) in the SGGHU are
  shown with green blobs.  The MSSM predictions are shown with red
  lines for $\tan \beta=10$ (thick line) and $\tan \beta=3$ (dashed).
  The NMSSM predictions are shown with blue grid lines, which indicate
  mixings between the SM-like and singlet like Higgs bosons of 10\%,
  20\% and 30\% from the right to the left.  }
\label{fig:vb}
\end{figure}

In Fig.~\ref{fig:vb}, the deviations in the Higgs boson coupling with
the weak gauge bosons $\kappa_V^{}$ and that with the bottom quark
$\kappa_b^{}$ from the SM predictions are plotted.  The predictions of
the three benchmark points (A), (B) and (C) in the SGGHU are shown
with green blobs.  The MSSM predictions are shown with red lines for
$\tan \beta=10$ (thick line) and $\tan \beta=3$ (dashed).  The NMSSM
predictions are shown with blue grid lines, which indicate mixings
between the SM-like and singlet like Higgs bosons of 10\%, 20\% and
30\% from the right to the left.  As is reported in
Ref.\cite{ILCHiggs}, the ILC with $\sqrt{s}=500~{\rm GeV}$ can reach
accuracy of 1.0\% (1.1\%) for the Higgs boson coupling with the
$Z$-boson (the $W$-boson).  Therefore, signatures different from the MSSM
and its variants are expected to be observed using $\kappa_V^{}$ at
the ILC.  Notice that the VEV of the triplet Higgs boson $v_\Delta^{}$
is small compared to those of the doublet Higgs bosons.  Therefore,
the mixing between the SM-like Higgs boson and the CP-even component
of the Higgs singlet dominates over that between the SM-like Higgs
boson and the triplet component.  In this sense, our model is similar
to the NMSSM.  It will be difficult to distinguish our model only from
these observables.

\begin{figure}[t]
\includegraphics[width=100mm]{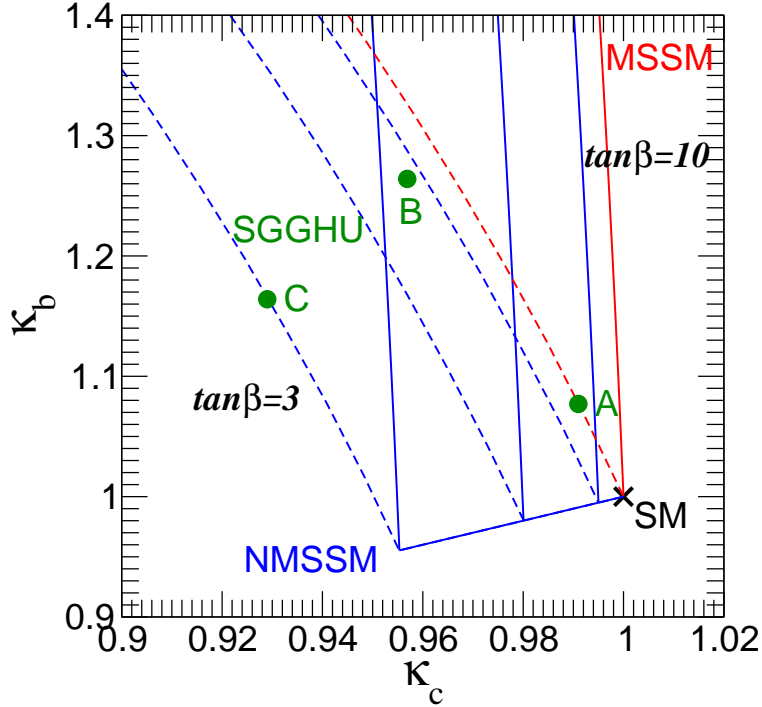}
\caption{\footnotesize The deviations in the Higgs boson coupling with
  the charm quark $\kappa_c^{}$ and that with the bottom quark
  $\kappa_b^{}$ from the SM predictions are plotted. See the caption
  of Fig.~\ref{fig:vb} for details.  }
\label{fig:cb}
\end{figure}

In Fig.~\ref{fig:cb}, the deviations in the Higgs boson coupling with the charm 
quark $\kappa_c^{}$ and that with the bottom quark $\kappa_b^{}$ from the SM
predictions are plotted.  As in Fig.~\ref{fig:vb}, the predictions of
the three benchmark points (A), (B) and (C) in the SGGHU are shown
with green blobs, and the MSSM and NMSSM predictions are shown with
red and blue lines, respectively.  In sharp contrast to the
$\kappa_V^{}$-$\kappa_b^{}$ relation, correlations between $\kappa_c^{}$ and
$\kappa_b^{}$ strongly depend on the value of $\tan \beta$.  For example,
the benchmark point (C) with $\tan \beta=3$ is not covered by the
NMSSM predictions with $\tan \beta=10$, and the deviation can be
measured at the ILC with $\sqrt{s}=500~{\rm GeV}$, which aims to
measure $\kappa_c^{}$ with accuracy of 2.8\%.  Independent $\tan \beta$
measurement using decay of the Higgs boson at the ILC
\cite{Gunion:2002ip,Kanemura:2013eja} will also play an important role
in discriminating models.  Although it will be difficult to completely
distinguish our model from the NMSSM from the precision measurements
of Higgs boson couplings, if the deviation pattern of the Higgs
couplings is found to be close to our benchmark points, there is a
fair possibility that the SGGHU is realized.  The ILC is absolutely
necessary for investigating the Higgs properties and distinguishing
particle physics models.

As for other Higgs boson couplings, the deviations of the Higgs boson coupling
with the photon are $0.94 < \kappa_\gamma^{} < 1.0$, and those of the
Higgs self-coupling $0.82 < \kappa_h^{} < 0.93$ for the benchmark
points we show.  To observe deviations in these observables from the
SM predictions one needs more precise measurements at the ILC with
$\sqrt{s}=1~{\rm TeV}$ \cite{ILCHiggs}.

\subsection{Additional Higgs bosons}

\begin{figure}[t]
\includegraphics[width=100mm]{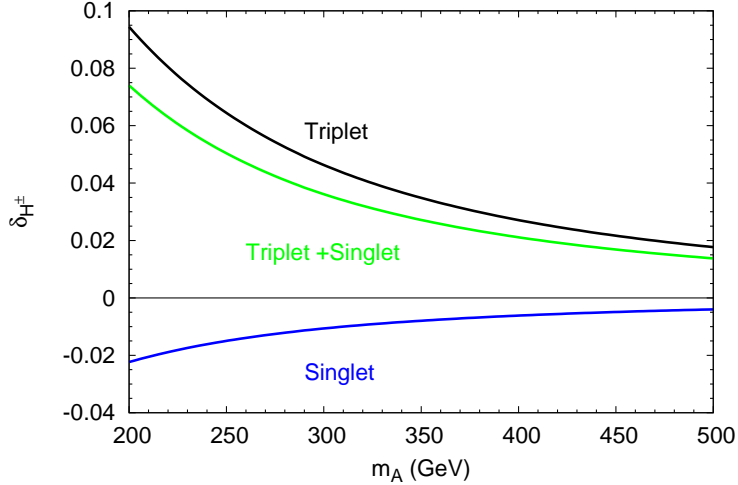}
\caption{\footnotesize The deviation parameter $\delta_{H^\pm}^{}$ of the
  MSSM-like charged Higgs boson mass $m_{H^\pm}^{}$ as a function of the
  MSSM-like CP-odd Higgs boson mass $m_A^{}$ in the large soft mass
  scenario. The black, blue and green lines correspond to triplet
  contribution, singlet contribution, sum of the singlet and triplet
  contributions, respectively. Here, we choose $\lambda_\Delta^{} = 1.1$
  and $\lambda_S^{} = 0.25$.}
\label{fig:dhpm_ma}
\end{figure}

Finally, we mention the additional MSSM-like Higgs bosons.  Since
four-point couplings in the Higgs sector are expressed in terms of
gauge couplings and $F$-term couplings in SUSY models, differences of
the masses of the MSSM-like Higgs bosons are also useful measures in
probing more fundamental physics. The MSSM-like charged Higgs boson mass
$m_{H^\pm}^{}$ is given by
\begin{align} 
  m_{H^\pm}^2
  &= m_{H^\pm}^2|_{\rm MSSM}^{} (1 +\delta_{H^\pm}^{})^2 \nonumber \\
  &\simeq m_A^2 +m_W^2 +\frac{1}{8}\lambda_\Delta^2 v^2
  -\frac{1}{2}\lambda_S^2v^2\, ,
\end{align}
where $\delta_{H^\pm}^{}$ is the deviation in $m_{H^\pm}^{}$ from the
MSSM and $m_A^{}$ is the MSSM-like CP-odd Higgs boson mass.  The sign
of the singlet contribution is opposite to the triplet one due to the
group theory.  From Eq.~(\ref{eq:couplings}), $m_{H^\pm}^{}$ becomes
large as compared to the MSSM.  We emphasize that these $\lambda_S^{}$
and $\lambda_\Delta^{}$ couplings are determined by the RGEs and a
larger $m_{H^\pm}^{}$ is a prediction in this model.  The charged
Higgs boson is always heavier than the CP-odd Higgs boson.  Since
$m_{H^\pm}^{}|_{\rm MSSM}^{}$ is the sum of $m_A^{}$ and $m_W^{}$, when
the CP-odd Higgs boson and the charged Higgs boson are discovered, we
can obtain $\delta_{H^\pm}^{}$ by measuring $m_A^{}$ and $m_{H^\pm}^{}$
precisely.  Fig.~\ref{fig:dhpm_ma} shows the deviation parameter
$\delta_{H^\pm}^{}$ of the MSSM-like charged Higgs boson mass $m_{H^\pm}^{}$
as a function of $m_A^{}$ in the large soft mass scenario.  The black,
blue and green lines correspond to triplet contribution, singlet
contribution, sum of the singlet and triplet contributions,
respectively. Here, we choose $\lambda_\Delta^{} = 1.1$ and $\lambda_S^{} =
0.25$.  The mass deviation is found to be $O(1)~\%$ - $O(10)~\%$ if
the mass scale of the MSSM-like Higgs bosons are below $500~{\rm
  GeV}$.  On the other hand, the deviation in the heavy CP-even Higgs
boson mass $m_H^{}$ from the MSSM prediction is less than $O(1)~\%$.
Since the charged Higgs boson mass can be determined with an accuracy
of a few percent at the LHC given such small masses \cite{HiggsWG}, we
can test our model.

When the masses of the triplet-like and singlet-like scalar bosons are
below $500~{\rm GeV}$, the ILC and CLIC have capability to directly
produce these new particles.  For example, the benchmark point (C)
gives mass spectrum of the Higgs sector particles shown in
Tab.~\ref{tb:spectrum}.  In this case, the mass of the lighter
triplet-like Higgs boson $\Delta^\pm$ is less than $500~{\rm GeV}$,
and we can probe $\Delta^\pm$ using the channel $e^+ e^- \rightarrow
\Delta^+ \Delta^- \rightarrow tb \bar{t} \bar{b}$, which proceeds via
the mixing between the MSSM-like and triplet-like charged Higgs bosons.

\begin{table}[t]
  \begin{tabular}{|c|c|c|}
    \hline
    CP-even & CP-odd & Charged \\ \hline \hline
    $122~{\rm GeV}$     & $-$                & $-$         \\ \hline
    $139~{\rm GeV}$     & $171~{\rm GeV}$    & $204~{\rm GeV}$     \\ \hline
    $370~{\rm GeV}$     & $304~{\rm GeV}$    & $496~{\rm GeV}$     \\ \hline
    $745~{\rm GeV}$     & $497~{\rm GeV}$    & $745~{\rm GeV}$     \\ 
    \hline
  \end{tabular}
  \caption{Mass spectrum of the Higgs scalars 
    for the benchmark point (C).} 
  \label{tb:spectrum}
\end{table}

\section{Discussion and Conclusion}

In this paper, we have investigated phenomenology of the Higgs sector
of the supersymmetric version of the grand gauge-Higgs unification
model, where the $SU(5)$ grand unified gauge symmetry is broken by the
Hosotani mechanism.  Our model provides a natural solution to the
doublet-triplet splitting problem thanks to the phase nature of the
Hosotani mechanism, and predicts existence of a light color octet, an
$SU(2)_L$ triplet and a neutral singlet chiral multiplets whose masses
are around the TeV scale.  Since the adjoint chiral multiplets are
originated from the GUT gauge multiplet, there are no trilinear
self-couplings among them and their couplings to MSSM fields are
unified to the SM gauge coupling constants at the GUT scale.
Therefore, our model is highly predictive.  We have performed RGE
analysis to obtain masses and coupling constants of the low-energy
effective theory of our model.  Although the mass scale of the color
octet chiral multiplet is found to be beyond reach of collider
experiments, the masses of the triplet and singlet multiplets can
remain as small as those of the MSSM Higgs doublets, and thus the
Higgs sector is extended by these new Higgs multiplets.

We have computed the SM-like Higgs boson mass including tree level and
one loop level contributions from the triplet and singlet couplings,
and shown benchmark points consistent with the LHC Higgs boson mass
measurements.  Based on the benchmark points, we have evaluated
deviations of couplings between the Higgs boson and SM particles from
the corresponding SM values, which are one of the main targets of the
future ILC project.  The deviations of the couplings from the SM
predictions turn out to be ${\cal O}(1)\%$ when the triplet and
singlet Higgs boson masses are below $\simeq 1~{\rm TeV}$.  Given such
small masses, we can distinguish our model, MSSM and NMSSM by comparing
patterns of the deviations of these new physics models.  As for
additional Higgs bosons, the mass gap between the MSSM-like charged
Higgs boson and the MSSM-like CP-odd Higgs boson differs from that of
the MSSM by ${\cal O}(1)\%$ - ${\cal O}(10)\%$ when their masses are
below $\simeq 500~{\rm GeV}$.  Such a deviation is within the scope of
the LHC.

Last but not least, the extension of the Higgs sector in SUSY models means
that the neutralino and chargino sectors are also extended.  For the
benchmark points we have shown, masses of the six neutralinos and
three charginos are all less than $500~{\rm GeV}$.  Collider
signatures of such additional neutralinos and charginos will be
discussed elsewhere.

We emphasize that our supersymmetric grand gauge-Higgs unification
model serves as a good example of grand unification that is testable
at future electron-positron colliders, and researches along this
strategy should be encouraged.

\begin{acknowledgments}
The work of S.K. was supported in part by Grant-in-Aid for Scientific Research, 
Japan Society for the Promotion of Science (JSPS)
and Ministry of Education, Culture, Sports, Science and Technology,
Nos. 22244031, 23104006 and 24340046.
The work of H.T. was supported in part by JSPS.

\end{acknowledgments}

\appendix
\section{Renormalization Group Equations}
\label{Sec:RGE}

In this appendix, we summarize the one loop RGEs between the SUSY scale
and the GUT scale for our model.  Here, for later reference, we give
them in a form which correctly includes the flavor structure though it
is less relevant to our analysis in this article.

%
\subsection{Notations}
%

In order to treat the flavor, it is convenient to use a notation different from the 
 one used in the main text for the $A$- and $B$-terms so that the corresponding SUSY 
 parameters are not extracted. 
To distinguish them, we append a bar on top of the $A$- and $B$-terms used in this 
 appendix. 
Namely, for example the $B$-term of the singlet is defined as $\bar B_S^{}=B_S^{}\mu_S^{}$. 

The flavor structure is expressed by using 3-by-3 matrices as usual. 
Here we use the character $Y$ for the Yukawa couplings with the flavor and thus 
 $Y$ is treated as a matrix, and the character $\lambda$ for those without 
 the flavor. 
The character $y$ denotes all the Yukawa coupling, $Y$ and $\lambda$, 
 symbolically.
A dot on a parameter $P(Q)$ is used for a partial derivative by the renormalization 
 scale $Q$ with a normalization factor:
 $(16 \pi^2)\partial P/\partial \ln (Q/Q_0)$ where $Q_0$ is an arbitrary 
 reference scale.

The superpotential we consider is
\begin{equation}
 W =  W^{\rm matter} +  W^{\rm Higgs}  +  W^{add}, 
\label{eq:superpotential}
\end{equation}
 with 
\begin{equation}
 W^{\rm matter} =  u Y_u q\cdot H_u - d Y_d q\cdot H_d -e Y_e \ell\cdot H_d, 
\label{eq:Wmatter}
\end{equation}
 $W^{\rm Higgs}$ is given in Eq.~(\ref{eq:WHiggs}) and 
\begin{eqnarray}
 W^{add} = 
   \bar L_j\cdot (\lambda_{L_j\Delta}\Delta - \lambda_{L_jS}S + \mu_{L_j})L_j
  +\bar U (\lambda_{UG}G - \frac43\lambda_{US}S+\mu_U)U
\nn\\
  +(2\lambda_{ES}S + \mu_E)\bar E E
  +\mu_G\tr(GG).
\end{eqnarray}
Here, $q$, $u$, $d$, $\ell$ and $e$ denote the MSSM matter chiral multiplets, 
 $Y_x$ ($x=u,d,e$) is a $3\times3$ matrix, $S$, $\Delta$ and $G$ are the adjoint 
 chiral multiplets and 
 $(\bar L_j,\ L_j)$ ($j=1,2$), $(\bar U,\ U)$ and $(\bar E,\ E)$
 are the additional vectorlike pairs introduced to recover the gauge coupling 
 unification%
\footnote{ 
In general, there exist the mixing terms between these vectorlike fields and 
 MSSM fields. 
The pattern of the mixing terms is highly model dependent while these 
mixings have little effects on the Higgs sector. 
Hence, we impose an additional 
 $\Z2$ symmetry that forbids such mixing terms to avoid unessential complication. 
}.

The SUSY-breaking soft terms contain the tadpole term of $S$, 
 aside from the usual soft mass squared terms, $A$-terms and $B$-terms:
\begin{equation}
 V_{\rm soft} =  
   V^{\rm matter}_{\rm soft} +  V^{\rm Higgs}_{\rm soft}  +  V^{add}_{\rm soft}, 
\label{eq:scalarpotential}
\end{equation}
 with 
\begin{eqnarray}
 V^{\rm matter}_{\rm soft} =  
   \s q^\dagger\s m^2_{q}\s q + \s u^T\s m^2_{u}\s u^* + \s d^T\s m^2_{d}\s d^*
  +\s \ell^\dagger\s m^2_{\ell}\s \ell + \s e^T\s m^2_{e}\s e^* \nn\\
  +\left[u \bar A_u q\cdot H_u -d \bar A_d q\cdot H_d -e \bar A_e \ell \cdot H_d 
  +{\rm h.c.}\right], 
\label{eq:Vmatter}
\end{eqnarray}
 $V^{\rm Higgs}_{\rm soft}$ is given in Eq.~(\ref{eq:VsoftHiggs}) and 
\begin{eqnarray}
 V^{add}_{\rm soft} = 
   \s m^2_{L_j}\abs{L_j}^2 + \s m^2_{\bar L_j}\abs{\bar L_j}^2
  +\s m^2_{U}\abs{U}^2 + \s m^2_{\bar U}\abs{\bar U}^2
  +\s m^2_{E}\abs{E}^2 + \s m^2_{\bar E}\abs{\bar E}^2
  +\s m^2_{G}\abs{G}^2 \nn\\
  +\left[\bar L_j\cdot (\bar A_{L_j\Delta}\Delta - \bar A_{L_jS}S + \bar B_{L_j})L_j
  +\bar U (\bar A_{UG}G - \frac43\bar A_{US}S + \bar B_U)U
\right.\nn\\\left.
  +(2\bar A_{ES}S + \bar B_E)\bar E E
  +\bar B_G\tr(GG) + {\rm h.c.}\right].
\end{eqnarray}
It is worthwhile to notice that the tadpole term of the scalar component of $S$ 
 in $V_{\rm soft}^{Higgs}$ 
 is generated even the tadpole term in the superpotential is forbidden, 
 while that of the $F$-component of $S$ can be removed by a field redefinition. 
Since the latter is generated by the loop corrections, 
 we have to do the field redefinition at each scale, 
 and the RGEs of the $B$-terms for the fields that couple to $S$ are affected.

%
\subsection{RGEs}
%

The formalism including some notations in this subsection is the one 
 in Ref.~\cite{BY}

\subsubsection{Gauge Couplings and Gaugino Masses}

The RGEs for the gauge couplings $g_i$ and the gaugino masses $M_i$ are given as
\begin{eqnarray}
 \dot{g}_i \, &=&  \, b_i \, g_i^3 , 
\nonumber\\ 
 \dot{M}_i    &=&  2b_i\, g_i^2 M_i, 
\label{eq:GaugGauginoRGEs}
\end{eqnarray} 
with the beta function coefficients $b_i$. 
In this model they are $b_i=(\frac{53}{5},5,1)$.

\subsubsection{Yukawa Couplings}

The RGEs for the Yukawa couplings:
\begin{eqnarray}
&\dot{Y}_u    &=
\gamma^T_{u} Y_u + Y_u \gamma_q + \gamma_{H_u}^{} Y_u ,
\nonumber\\
&\dot{Y}_d    &=
\gamma^T_{d} Y_d + Y_d \gamma_q + \gamma_{H_d}^{} Y_d ,
\nonumber\\
&\dot{Y}_e    &=
\gamma^T_{e} Y_e + Y_e \gamma_\ell + \gamma_{H_d}^{} Y_e ,
\nn\\
&\dot{\lambda}_\Delta^{}    &=
(\gamma_{H_u}^{}+\gamma_{H_d}^{}+\gamma_{\Delta}) \lambda_\Delta^{} ,
\nn\\
&\dot{\lambda}_S^{}    &=
(\gamma_{H_u}^{}+\gamma_{H_d}^{}+\gamma_{S}) \lambda_S^{} ,
\nn\\
&\dot{\lambda}_{L_j\Delta}    &=
(\gamma_{L_j}+\gamma_{\bar L_j}+\gamma_{\Delta}) \lambda_{L_j\Delta} ,
\nn\\
&\dot{\lambda}_{L_jS}    &=
(\gamma_{L_j}+\gamma_{\bar L_j}+\gamma_{S}) \lambda_{L_jS} ,
\nn\\
&\dot{\lambda}_{UG}    &=
(\gamma_{U}+\gamma_{\bar U}+\gamma_{G}) \lambda_{UG} ,
\nn\\
&\dot{\lambda}_{US}    &=
(\gamma_{U}+\gamma_{\bar U}+\gamma_{S}) \lambda_{US} ,
\nn\\
&\dot{\lambda}_{ES}    &=
(\gamma_{E}+\gamma_{\bar E}+\gamma_{S}) \lambda_{ES} ,
\label{eq:RGEsYukawa}
\end{eqnarray}
 with the anomalous dimensions $\gamma$, among which
 $\gamma_f$ ($f=q,u,d,\ell,e$) is a $3\times3$ matrix.
\begin{eqnarray}
&\gamma_q                          & =  
 -2\left(\displaystyle{\frac{4}{3}}g^2_3 
                      +\frac{3}{4}g^2_2 +\frac{1}{60}g^2_1 
   \right) \!{\mathop{\bf 1}}
 +Y_u^\dagger Y_u +Y_d^\dagger Y_d , 
\nonumber\\
&\gamma_{u}                        & =  
 -2\left(\displaystyle{\frac{4}{3}}g^2_3
                      +\frac{4}{15}g^2_1 \right) \!{\mathop{\bf 1}}
 + 2 Y_u^\ast Y_u^T ,
\nonumber\\
&\gamma_{d}                        & =  
 -2\left(\displaystyle{\frac{4}{3}}g^2_3
                      +\frac{1}{15}g^2_1 \right) \!{\mathop{\bf 1}}
 + 2 Y_d^\ast Y_d^T ,
\nonumber\\
&\gamma_\ell                          & =  
 -2\left(\displaystyle{\frac{3}{4}}g^2_2
                      +\frac{3}{20}g^2_1 \right) \!{\mathop{\bf 1}}  
 +Y_e^\dagger Y_e ,
\nonumber\\
&\gamma_{e}                        & =  
 -2\left(\displaystyle{\frac{3}{5}}g^2_1 \right) \!{\mathop{\bf 1}}  
 + 2 Y_e^\ast Y_e^T , 
\nonumber\\[1.001ex]
&\gamma_{H_u}^{}                      & =  
 -2\left(\displaystyle{\frac{3}{4}}g^2_2 +\frac{3}{20}g^2_1 \right)
 + {\rm Tr} \left(3 Y_u^\dagger Y_u\right)+\delta\gamma_H,
\nonumber\\
&\gamma_{H_d}^{}                      & =  
 -2\left(\displaystyle{\frac{3}{4}}g^2_2 +\frac{3}{20}g^2_1 \right)
 + {\rm Tr} \left(3 Y_d^\dagger Y_d + Y_e^\dagger Y_e\right)+\delta\gamma_H,
\label{eq:gammaMSSM}
\end{eqnarray}
\begin{eqnarray}
&\delta\gamma_H                    &= 
  \abs{\lambda_S^{}}^2 + \frac34\abs{\lambda_\Delta^{}}^2,
\nonumber\\[1.001ex]
&\gamma_S^{}                          &= 
  2\abs{\lambda_S^{}}^2 + \sum_j2\abs{\lambda_{L_jS}}^2
 +\frac{16}3\abs{\lambda_{US}}^2 + 4\abs{\lambda_{ES}}^2,
\nonumber\\
&\gamma_{\Delta}                   & =  
 -2\left(\displaystyle{2}g^2_2 \right)
 +\frac12\abs{\lambda_\Delta^{}}^2 + \sum_j\frac12\abs{\lambda_{L_j\Delta}}^2,
\nonumber\\
&\gamma_{G}                        & =  
 -2\left(\displaystyle{3}g^2_3 \right)
 +\frac12\abs{\lambda_{UG}}^2,
\nonumber\\
&\gamma_{L_j} = \gamma_{\bar L_j}  & =  
 -2\left(\displaystyle{\frac{3}{4}}g^2_2 +\frac{3}{20}g^2_1 \right)
 + \abs{\lambda_{L_jS}}^2 + \frac34\abs{\lambda_{L_j\Delta}}^2,
\nonumber\\
&\gamma_{U} = \gamma_{\bar U}      & =  
 -2\left(\displaystyle{\frac{4}{3}}g^2_3
                      +\frac{4}{15}g^2_1 \right)
 + \frac{16}9\abs{\lambda_{US}}^2 + \frac43\abs{\lambda_{UG}}^2,
\nonumber\\
&\gamma_{E} = \gamma_{\bar E}      & =  
 -2\left(\frac{3}{5}g^2_1 \right)
 + 4\abs{\lambda_{ES}}^2.
\label{eq:gammaAdd}
\end{eqnarray}

\subsubsection{$\mu$-terms}

Similarly, the supersymmetric mass terms evolve as
\begin{eqnarray}
&\dot{\mu}           &=  \left( \gamma_{H_u}^{} + \gamma_{H_d}^{} \right)\mu ,
\nonumber\\
&\dot{\mu_S^{}}         &=  2 \gamma_{S}\mu_S^{} ,
\nonumber\\
&\dot{\mu_\Delta^{}}    &=  2 \gamma_{\Delta}\mu_\Delta^{} ,
\nonumber\\
&\dot{\mu_G}         &=  2 \gamma_{G}\mu_G ,
\nonumber\\
&\dot{\mu_{L_j}}     &=  \left( \gamma_{\bar L_j} + \gamma_{L_j} \right)\mu_{L_j} ,
\nonumber\\
&\dot{\mu_U}         &=  \left( \gamma_{\bar U} + \gamma_{U} \right)\mu_{U} ,
\nonumber\\
&\dot{\mu_E}         &=  \left( \gamma_{\bar E} + \gamma_{E} \right)\mu_{E} .
\label{eq:RGEsmu}
\end{eqnarray}

\subsubsection{$A$-terms}

The RGEs for the $A$-terms can be derived from those for the corresponding 
 Yukawa couplings given in Eq.~\eqref{eq:RGEsYukawa} which are functions of the Yukawa 
 couplings $y$ and the anomalous dimensions $\gamma$, $\dot y(y,\gamma)$, by
\begin{equation}
 \dot{\bar A}_x = 
 \dot{y}_x \vert_{y \to \bar A} 
+\dot{y}_x \vert_{\gamma \to \tilde{\gamma}} 
=\dot{y}_x (\bar A,\gamma)
+\dot{y}_x (y,\tilde\gamma).
\label{eq:RGEsA}
\end{equation}
Here, the quantities $\tilde{\gamma}_f$ can be built from the corresponding anomalous
 dimensions in Eqs.~\eqref{eq:gammaMSSM} and \eqref{eq:gammaAdd} with the replacements:
\begin{equation}
 g_i^2           \to  - 2 g_i^2 M_i,        \hspace*{1truecm}
 y_x^\dagger y_x \to    2 y_x^\dagger \bar A_x,  \hspace*{1truecm}
 y_x^\ast y_x^T  \to    2 y_x^\ast \bar A_x^T.  
\label{eq:gamma2sgamma}
\end{equation}
%

\subsubsection{$B$-terms}

Similarly, the RGEs for the $B$-terms are obtained from those for the corresponding 
 $\mu$-terms in Eq.~\eqref{eq:RGEsmu}, but with further contribution 
 due to the field redefinition of $S$ as
\begin{equation}
 \dot{\bar B}_f = 
 \dot{\mu}_f \vert_{\mu \to \bar B} 
+\dot{\mu}_f \vert_{\gamma \to \tilde{\gamma}}
+\lambda_{fS}J_S^{}.
\label{eq:RGEsB}
\end{equation}
Here, $\lambda_{fS}$ is the Yukawa couplings among the relevant vectorlike pair and 
 the singlet $S$, and
 the quantity $J_S^{}$ is built from $\gamma_S^{}$ in Eq.~\eqref{eq:gammaAdd} 
 with the replacement
\begin{equation}
 \abs{\lambda_{fS}}^2\to 2\lambda_{fS}^* \bar B_f.  
\label{eq:gammaS2tau}
\end{equation}

\subsubsection{Scalar Tadpole}

The RGE for the scalar tadpole of $S$ is written by
\begin{equation}
\dot\xi=\gamma_S^{}\xi + \s J_S^* + J_S^{}\mu_S^{},
\label{eq:RGEsxi}
\end{equation}
with $\s J_S^{}$ built from $\gamma_S^{}$ in Eq.~\eqref{eq:gammaAdd} 
 with the replacement
\begin{equation}
 \abs{\lambda_{fS}}^2
 \to 2\left(\lambda_{fS}^\dagger\left(\s m^2_f+\s m^2_{\bar f}\right)\mu_f
           + \bar A_{fS}^* \bar B_f\right),  
\label{eq:gammaS2stau}
\end{equation}
 where $\s{m}^2_f$ and $\s{m}^2_{\bar f}$ are the masses squared of
 the relevant vectorlike pair.

\subsection{Scalar Soft Squared Masses}

It is convenient to define a function ${\cal F}$:
\begin{eqnarray}
{\cal F}(y_x^\dagger,f_1,f_2,f_3,
         \bar A_x^\dagger)
&=& 
   y_x^\dagger y_x \widetilde{m}^2_{f_1}
+ \widetilde{m}^2_{f_1} y_x^\dagger y_x
\nonumber \\ 
&&
+ 2\left(   y_x^\dagger \widetilde{m}^{2\,\ast}_{f_2} y_x +\!
            \widetilde{m}^2_{f_3} y_x^\dagger y_x         +\! 
            \bar A_x^\dagger \bar A_x              \right) ,\quad
\label{eq:calFuncDEF}
\end{eqnarray}
 where $y_x$ is any of the Yukawa couplings in the superpotential,
 $\s{m}^2_{f_1}$ is the mass squared for which the RGE in question
 is derived,
 $\s{m}^2_{f_2}$ and $\s{m}^2_{f_3}$ are the masses squared of
 the particles exchanged in the loops that induce the RGE. 
Since the order of $f_1,f_2,f_3$ is not important when $y_x$ is some of $\lambda$, 
 the order may be changed below.
 
Using the function $\cal F$, the RGEs for the soft squared masses are written as 
\begin{eqnarray}
&\dot{\widetilde{m}}^2_q  \,      
 &=
 -8\left(\frac{4}{3}g^2_3  M_3^2 
        +\frac{3}{4}g^2_2  M_2^2
        +\frac{1}{60}g^2_1 M_1^2  \right) \!{\mathop{\bf 1}}
 +{\cal F}(Y_u^\dagger,{q},{u},H_u,\bar A_u^\dagger)
 +{\cal F}(Y_d^\dagger,{q},{d},H_d,\bar A_d^\dagger),
\nonumber\\
&\dot{\widetilde{m}}_{u}^2      
&=
 -8\left(\frac{4}{3}g^2_3  M_3^2 
        +\frac{4}{15}g^2_1 M_1^2  \right) \!{\mathop{\bf 1}}
 +2{\cal F}{(Y_u,{u},{q},H_u,\bar A_u)} ,
\nonumber\\
&\dot{\widetilde{m}}_{d}^2     
&=
 -8\left(\frac{4}{3}g^2_3  M_3^2 
        +\frac{1}{15}g^2_1 M_1^2  \right) \!{\mathop{\bf 1}}
 +2{\cal F}{(Y_d,{d},{q},H_d,\bar A_d)} ,
\nonumber\\
&\dot{\widetilde{m}}_\ell^2         
&=
 -8\left(
        \frac{3}{4}g^2_2  M_2^2
        +\frac{3}{20}g^2_1 M_1^2  \right) \!{\mathop{\bf 1}}
 +{\cal F}{(Y_e^\dagger,{\ell},{e},H_d,\bar A_e^\dagger)},
\nonumber\\
&\dot{\widetilde{m}}_{e}^2      
&=
 -8\left(
        \frac{3}{5}g^2_1 M_1^2  \right) \!{\mathop{\bf 1}}
 +2{\cal F}{(Y_e,e,\ell,H_d,\bar A_e)} ,
\label{eq:RGEsSoftMassMatter}
\end{eqnarray}
%
%
\begin{eqnarray}
&\dot{\widetilde{m}}_{H_d}^2   
& =
 -8\left(
        \frac{3}{4}g^2_2  M_2^2
        +\frac{3}{20}g^2_1 M_1^2  \right)
\nonumber\\[1.1ex]&&\qquad
 +3{\rm Tr}{\cal F}{(Y_d^\dagger,H_d,d,q,\bar A_d^\dagger)} 
 + {\rm Tr}{\cal F}{(Y_e^\dagger,H_d,e,\ell,\bar A_e^\dagger)}
 + \delta\dot{\widetilde{m}}_{H}^2,
\nonumber\\[1.1ex]
&\dot{\widetilde{m}}_{H_u}^2   
& =
 -8\left(
        \frac{3}{4}g^2_2  M_2^2
        +\frac{3}{20}g^2_1 M_1^2  \right)
 +3{\rm Tr}{\cal F}{(Y_u^\dagger,H_u,u,q,\bar A_u^\dagger)}
 + \delta\dot{\widetilde{m}}_{H}^2,
\nonumber\\[1.1ex]
&\delta\dot{\widetilde{m}}_{H}^2 
& = 
  {\cal F}{(\lambda_S^{},H_u,H_d,S,\bar A_S^{})}
 +\frac34{\cal F}{(\lambda_\Delta^{},H_u,H_d,\Delta,\bar A_\Delta^{})},
\label{eq:RGEsSoftMassHiggs}
\end{eqnarray}
\begin{eqnarray}
&\dot{\widetilde{m}}_{S}^2   
& =
 2{\cal F}(\lambda_S^{},S,H_u,H_d,\bar A_S^{})
 +\sum_j2{\cal F}(\lambda_{L_jS},S,L_j,\bar L_j,\bar A_{L_jS})
\nonumber\\[1.1ex]&&\quad
 +\frac{16}3{\cal F}(\lambda_{US},S,U,\bar U,\bar A_{US})
 +4{\cal F}(\lambda_{ES},S,E,\bar E,\bar A_{ES}),
\nonumber\\[1.1ex]
&\dot{\widetilde{m}}_{\Delta}^2   
& =
 -8\left(
        2g^2_2  M_2^2\right)
 +\frac12{\cal F}(\lambda_\Delta^{},\Delta,H_u,H_d,\bar A_\Delta^{})
 +\sum_j\frac12{\cal F}(\lambda_{L_j\Delta},\Delta,L_j,\bar L_j,\bar A_{L_j\Delta}),
\nonumber\\[1.1ex]
&\dot{\widetilde{m}}_{G}^2   
& =
 -8\left(3g^2_3  M_3^2\right) 
 +\frac12{\cal F}(\lambda_{UG},G,U,\bar U,\bar A_{UG}),
\nonumber\\[1.1ex]
&\dot{\widetilde{m}}_{L_j}^2 = \dot{\widetilde{m}}_{\bar L_j}^2   
& =
 -8\left(
         \frac34g^2_2  M_2^2
        +\frac{3}{20}g^2_1 M_1^2  \right)
\nonumber\\[1.1ex]&&\qquad
 +{\cal F}(\lambda_{L_jS},L_j,\bar L_j,S,\bar A_{L_jS})
 +\frac34{\cal F}(\lambda_{L_j\Delta},L_j,\bar L_j,\Delta,\bar A_{L_j\Delta}),
\nonumber\\[1.1ex]
&\dot{\widetilde{m}}_{U}^2 = \dot{\widetilde{m}}_{\bar U}^2   
& =
 -8\left(\frac43g^2_3  M_3^2 
        +\frac{4}{15}g^2_1 M_1^2  \right)
\nonumber\\[1.1ex]&&\qquad
 +\frac{16}9{\cal F}(\lambda_{US},U,\bar U,S,\bar A_{US})
 +\frac43{\cal F}(\lambda_{UG},U,\bar U,G,\bar A_{UG}),
\nonumber\\[1.1ex]
&\dot{\widetilde{m}}_{E}^2 = \dot{\widetilde{m}}_{\bar E}^2   
& =
 -8\left(
        \frac{3}{5}g^2_1 M_1^2  \right)
 +4{\cal F}(\lambda_{ES},E,\bar E,S,\bar A_{ES}).
\label{eq:RGEsSoftAdd}
\end{eqnarray}
Here, we omit the corrections due to the $D$-term interactions of the hypercharge, 
\[
 \frac65 y_fg_1^2 \tr(y_{f'}\s m_{f'}^2), 
\]
 since they vanish when we take (semi-)universal boundary condition.

%
%

%
\subsection{Input Parameters at the GUT Scale}
%

In this analysis, we assume certain universality among the SUSY breaking parameters at the GUT scale, 
 for simplicity. 
The gaugino masses $M_{1/2}$ should be common since our model is a kind of the grand unified theory. 
The $A$-terms and soft squared masses for the MSSM matter (quark and lepton) multiplets are set common 
 value as $A_0$ (with the notation in the main text) and $\tilde m^2_0$, respectively. 
We treat the soft squared masses for the doublet Higgs multiplets, 
 $\tilde m^2_{H_u}$ and $\tilde m^2_{H_d}$, 
 as independent parameters not necessarily equal to $\tilde m^2_0$. 
Their $\mu$-term and the $B$-term are also free parameters.

Since the adjoint chiral multiplets originate from the unified gauge
multiplet, their parameters should be common at the cutoff scale, and
one of two real scalar field in each adjoint multiplets should be
massless while its SUSY partners, the other scalar and the Majorana
fermion, can be massive.  This fact allows us to introduce a
$\mu$-parameter, $\mu_\Sigma$, for the mass of the fermionic
component, a soft squared mass, $\tilde m^2_{\Sigma}$, and a
$B$-parameter, $\bar B_\Sigma$, for the adjoint multiplets that
satisfy a relation $\abs{\bar B_\Sigma}=\abs{\tilde
  m^2_{\Sigma}+\mu_{\Sigma}^2}$.  In addition, the $A$-terms for the
adjoint Yukawa couplings are forbidden.  Although the scalar tadpole
term for the massive scalar in the singlet multiplet is not forbidden,
we do not introduce it for simplicity.

As for the additional vectorlike pairs, we take common parameters for the pairs $(\bar U, U)$ and 
 $(\bar E,E)$ since they are assumed to be unified into a single {\bf 10} multiplet.
The parameters for the two pairs $(\bar L_i, L_i)$ could depend on the "flavor" $i$, but we take 
 common parameters here.
For these, we have the $\mu$-parameters, $\mu_{10}$ and $\mu_{5}$, $B$-parameters, $B_{10}$ and $B_{5}$,
 and the soft squared masses, $\tilde m^2_{10}$ and $\tilde m^2_{5}$.

In summary, our parameters at the GUT scale are one gaugino mass, one $A$-parameter, 
 four $\mu$-parameters, four $B$-parameters and six soft squared masses, 
 with one condition for a massless adjoint scalar. 

Among these parameters, the $\mu$ and $B$-parameters of the additional vector-like 
 pairs do not affect the running of the parameters in the Higgs sector, and we fix 
 them as $\mu_{10}=-20$ TeV and $\mu_{5}=5$ TeV so that they are decoupled from the 
 sub-TeV physics, and $B_{10}=B_{5}=0$ for simplicity.

\section{Masses and Mixings of the Higgs Bosons}
\label{Sec:Higgsmass}

Here we detail computations of the masses and mixings of the Higgs
bosons.  

\subsection{Higgs Potential}

The Higgs superpotential and soft SUSY breaking terms are
given by Eqs.~(\ref{eq:WHiggs}) and (\ref{eq:VsoftHiggs}),
respectively.  The scalar components of the MSSM Higgs superfields are
expanded around their VEVs as
\begin{eqnarray}
  H_u = \left( 
    \begin{array}{c}
      h_u^+ \\ v_u/\sqrt{2} + h_u^0 
    \end{array}
    \right)\, ,    \quad
  H_d = \left( 
    \begin{array}{c}
      v_d/\sqrt{2} + h_d^0 \\ h_d^-
    \end{array}
    \right)\, ,
\end{eqnarray}
and those of triplet and singlet as
\begin{eqnarray}
  \Delta = 
  \left(
    \begin{array}{cc}
      (v_\Delta^{} + \Delta^0)/2 & \bar{\Delta}^+/\sqrt{2} \\
      \Delta^-/\sqrt{2} & - (v_\Delta^{} + \Delta^0)/2 
    \end{array}
    \right)\, , \quad S = (v_S^{} + S^0)/\sqrt{2}\, .
\end{eqnarray}
The minimum of the tree-level Higgs potential is obtained by using 
the tadpole conditions,
\begin{eqnarray}
  \frac{\partial V}{\partial v_u}
  & = & 
  v_d \left( \widetilde{m}_{H_d}^2 + \mu_{\rm eff}^2 + \frac{\lambda_\Delta^2 v_u^2}{8}
    + \frac{\lambda_S^2 v_u^2}{2}
    + \frac{g_2^2 + g_1^2}{8}(v_d^2 - v_u^2) \right) - v_u \hat{m}_3^2 = 0\, ,
  \nonumber \\
  \frac{\partial V}{\partial v_d}
  & = & 
  v_u \left( \widetilde{m}_{H_u}^2 + \mu_{\rm eff}^2 + \frac{\lambda_\Delta^2 v_d^2}{8}
    + \frac{\lambda_S^2 v_d^2}{2}
    + \frac{g_2^2 + g_1^2}{8}(v_u^2 - v_d^2) \right) - v_d \hat{m}_3^2 = 0\, ,
  \nonumber \\
  \frac{\partial V}{\partial v_\Delta^{}}
  &=& 
  v_\Delta^{} (\widetilde{m}_\Delta^2 + m_{3\Delta}^2 + \mu_\Delta^2 )
  - \frac{\lambda_\Delta^{}}{2\sqrt{2}} \left[
    (\mu_\Delta^{} + A_\Delta^{})v_u v_d - \mu_{\rm eff} (v_u^2 + v_d^2) \right] = 0 \, ,
  \nonumber \\
  \frac{\partial V}{\partial v_S^{}}
  & = & v_S^{} (\widetilde{m}_S^2 + m_{3S}^2 + \mu_S^2 ) 
  - \frac{\lambda_S^{}}{\sqrt{2}}
  [(\mu_S^{} + A_S^{}) v_d v_u - \mu_{\rm eff}(v_d^2 + v_u^2)]
  + \sqrt{2} \xi
  = 0\, ,
\end{eqnarray}
where we have replaced $\mu$ and $m_3^2$ by
\begin{eqnarray}
  \mu_{\rm eff} & \equiv & \mu + \frac{\lambda_\Delta^{} v_\Delta^{}}{2\sqrt{2}}
  + \frac{\lambda_S^{} v_S^{}}{\sqrt{2}}
\, ,
  \nonumber \\
  \hat{m}_3^2 & \equiv & m_3^2
  + \frac{\lambda_\Delta^{} v_\Delta^{}}{2\sqrt{2}} (\mu_\Delta^{} + A_\Delta^{})
  + \frac{\lambda_S^{} v_S^{}}{\sqrt{2}} (\mu_S^{} + A_S^{})\, ,
\end{eqnarray}
respectively.  These parameters play roles similar to $\mu$ and
$m_3^2$ in the MSSM, and are derived from the tadpole conditions as
\begin{eqnarray}
  \mu_{\rm eff}^2 &=& - \frac{g_2^2+g_1^2}{8}v^2
  - \frac{t_\beta^2}{t_\beta^2-1}\widetilde{m}_{H_u}^2
  + \frac{1}{t_\beta^2-1}\widetilde{m}_{H_d}^2\, ,
  \nonumber \\
  \frac{\hat{m}_3^2}{c_\beta s_\beta} & = &
  \widetilde{m}_{H_u}^2+\widetilde{m}_{H_d}^2+2 \mu_{\rm eff}^2 
  + \frac{\lambda_\Delta^2 v^2}{8} + \frac{\lambda_S^2 v^2}{2}\, ,
\end{eqnarray}
where we have defined
\begin{eqnarray}
  \tan \beta = \frac{v_u}{v_d}\, , \quad v^2 = v_u^2 + v_d^2\, ,
\end{eqnarray}
and used the abbreviations $s_\beta = \sin \beta$, $c_\beta = \cos \beta $
and $t_\beta = \tan \beta$.

Data of electroweak precision measurements show that the rho
parameter is very close to one: the VEV of the neutral component of
the Higgs triplet field is much smaller than $v$.
Therefore, the mass matrices of the Higgs bosons can be expanded with
respect to $v_\Delta^{}/v$.  Hereafter, we keep only the leading term
in each Higgs mass matrix taking the limit of $v_\Delta^{}/v \to 0$.

\subsection{Higgs Mass Matrices}

In the basis of $w^-_i = (h_d^-, h_u^-, \bar{\Delta}^-, \Delta^- )$,
the mass squared matrix of the charged Higgs bosons is given by
\begin{eqnarray}
{\footnotesize
  {\cal M}_\pm^2 = 
\left( 
    \begin{array}{cccc}
      \hat{M}_C^2 s_\beta^2 & 
      \hat{M}_C^2 s_\beta c_\beta & 
      - \frac{\lambda_\Delta^{}}{2}v s_\beta (\mu_{\rm eff} t_\beta - \mu_\Delta^{}) &
      \frac{\lambda_\Delta^{}}{2} v s_\beta (\mu_{\rm eff}/t_\beta - \mu_\Delta^{}) 
      \\
      \cdots &
      \hat{M}_C^2 c_\beta^2 & 
      - \frac{\lambda_\Delta^{}}{2}v c_\beta (\mu_{\rm eff} t_\beta - \mu_\Delta^{})
      & 
      \frac{\lambda_\Delta^{}}{2}v c_\beta (\mu_{\rm eff}/ t_\beta - \mu_\Delta^{})
      \\
      \cdots & \cdots & 
       \mu_\Delta^2 + \widetilde{m}_\Delta^2 + \frac{g_2^2}{4} v^2 c_{2\beta}
      + \frac{\lambda_\Delta^2}{4}v^2 s_\beta^2 &
      m^2_{3 \Delta} 
      \\
      \cdots & \cdots & \cdots & 
      \mu_\Delta^2 + \widetilde{m}_\Delta^2 - \frac{g_2^2}{4}v^2 c_{2\beta}
      + \frac{\lambda_\Delta^2}{4}v^2 c_\beta^2 
   \end{array}
\right)\, ,
}
\end{eqnarray}
where 
\begin{eqnarray}
  \hat{M}_C^2 = \frac{\hat{m}_3^2}{s_\beta c_\beta} + 
      \left( \frac{g_2^2}{4} + \frac{\lambda_\Delta^2}{8} - \frac{\lambda_S^2}{2}
        \right)
      v^2 \, .
\end{eqnarray}
The mass eigenstates of the charged Higgs bosons $h_i^-$ are obtained by 
a unitary matrix $U^C$ as
\begin{eqnarray}
  h^-_i = U_{ij}^C w^-_j\, .
\end{eqnarray}
In the limit of heavy triplet and singlet components, the mass squared
of the MSSM-like charged Higgs boson is approximately given by
\begin{eqnarray}
  m_{H^{\pm}}^2 \simeq \hat{M}_C^2 = \frac{\hat{m}^2_3}{s_\beta c_\beta }
 + \left( \frac{g_2^2}{4} + \frac{\lambda_\Delta^2}{8} 
      - \frac{\lambda_S^2}{2} \right) v^2\, .
\end{eqnarray}

In the basis of $P_i = ({\rm Im}(h_d^0)/\sqrt{2}, {\rm Im}(h_u^0)/\sqrt{2}, {\rm Im}(S^0)/\sqrt{2}, {\rm Im}(\Delta^0)/\sqrt{2})$, the mass squared matrix of 
the CP-odd Higgs bosons is given by
\begin{eqnarray}
  {\cal M}_A^{2} = \left( 
    \begin{array}{cccc}
      \hat{m}_3^2 t_\beta & 
      \hat{m}_3^2  & 
      - \frac{\lambda_S^{}}{\sqrt{2}}v s_\beta  (\mu_S^{} - A_S^{}) &
      \frac{\lambda_\Delta^{}}{2\sqrt{2}}\frac{v}{c_\beta}
      (\mu_{\rm eff} - \mu_\Delta^{} s_{2\beta})
      \\
      \cdots & \hat{m}_3^2/t_\beta &  
      - \frac{\lambda_S^{}}{\sqrt{2}}v c_\beta  (\mu_S^{} - A_S^{})  &
      \frac{\lambda_\Delta^{}}{2\sqrt{2}}\frac{v}{s_\beta}
      (\mu_{\rm eff} - \mu_\Delta^{} s_{2\beta}) 
      \\
      \cdots & \cdots &
      \mu^2_S + \widetilde{m}^2_S - m_{3S}^2 + \frac{\lambda_S^2}{2} v^2 &
      \frac{\lambda_\Delta^{} \lambda_S^{}}{4}v^2
      \\
      \cdots & \cdots & \cdots &  
      \mu^2_\Delta + \widetilde{m}^2_\Delta - m_{3\Delta}^2 + 
      \frac{\lambda_\Delta^2}{8} v^2
      \\
    \end{array}
\right)\, .
\end{eqnarray}
The mass eigenstates of the CP-odd Higgs bosons $a_i$ are obtained
by an orthogonal matrix $R^P$ as
\begin{eqnarray}
  a_i = R^P_{ij} P_j\, .
\end{eqnarray}
In the limit of heavy triplet and singlet components, the mass squared
of the MSSM-like CP-odd Higgs boson by
\begin{eqnarray}
  m_A^2 \simeq \frac{\hat{m}_3^2}{s_\beta c_\beta}\, .
\end{eqnarray}
Therefore, the mass squared difference between the MSSM-like charged and
CP-odd Higgs bosons is
\begin{eqnarray}
  m_{H^{\pm}}^2 - m_A^2 \simeq \left( \frac{g_2^2}{4} + \frac{\lambda_\Delta^2}{8} 
      - \frac{\lambda_S^2}{2} \right) v^2\, .
\end{eqnarray}
Given the above charged (CP-odd) Higgs boson mass matrix, 
the eigenstate whose mass eigenvalue vanishes corresponds to 
the Nambu-Goldstone boson absorbed by the $W$-($Z$-) boson.

In the basis of $S_i=({\rm Re}(h_d^0)/\sqrt{2}, {\rm Re}(h_u^0)/\sqrt{2}, {\rm Re}(S^0)/\sqrt{2}), {\rm Re}(\Delta^0)/\sqrt{2})$, the mass squared matrix
of the CP-even Higgs bosons is given by
\begin{eqnarray}
{\footnotesize
  {\cal M}_S^{2} = 
\left( 
    \begin{array}{cccc}
      \hat{m}_3^2 t_\beta + m_Z^2 c_\beta^2& 
      - \hat{m}_3^2 - \hat{M}^2 s_\beta c_\beta&
      \frac{\lambda_S^{}}{\sqrt{2}}v (2\mu_{\rm eff}c_\beta -(\mu_S^{} + A_S^{}) s_\beta) &
      \frac{\lambda_\Delta^{}}{2\sqrt{2}} \mu_{\rm eff}v\frac{c_{2\beta}}{c_\beta}
      \\
      \cdots &
      \frac{\hat{m}_3^2}{t_\beta} + m_Z^2 s_\beta^2 &   
      \frac{\lambda_S^{}}{\sqrt{2}}v (2\mu_{\rm eff}s_\beta -(\mu_S^{} + A_S^{}) c_\beta) &
      - \frac{\lambda_\Delta^{}}{2\sqrt{2}} \mu_{\rm eff}v\frac{c_{2\beta}}{s_\beta}
      \\
      \cdots & \cdots & 
      \mu^2_S + \widetilde{m}^2_S + m_{3S}^2 + \frac{\lambda_S^2}{2} v^2 &
      \frac{\lambda_\Delta^{} \lambda_S^{}}{4}v^2
      \\
      \cdots & \cdots & \cdots &  
      \mu^2_\Delta + \widetilde{m}^2_\Delta + m_{3\Delta}^2 + 
      \frac{\lambda_\Delta^2}{8} v^2
      \\
    \end{array}
\right)\, ,
} \nonumber
\\
\end{eqnarray}
where
\begin{eqnarray}
  \hat{M}^2 = \left( \frac{1}{4} g_1^2 + \frac{1}{4} g_2^2 
      - \frac{1}{4} \lambda_\Delta^2 - \lambda_S^2 \right)v^2\, .
\end{eqnarray}
The mass eigenstates of the CP-even Higgs bosons $h_i$ are obtained
by an orthogonal matrix $R^S$ as
\begin{eqnarray}
  h_i = R^S_{ij} S_j\, .
\end{eqnarray}
At the tree-level, the mass eigenvalues of the MSSM-like CP-even Higgs
bosons are approximately given by
\begin{eqnarray}
  m_h^2
  & \simeq & 
  m_Z^2 \cos^2{2\beta}
  + \left( \frac{\lambda_\Delta^2}{8} + \frac{\lambda_S^2}{2} \right)
  v^2 \sin^2 {2\beta}\, ,
  \nonumber \\
  m_H^2 & \simeq & 
  \frac{\hat{m}^2_3}{s_\beta c_\beta}
  + m_Z^2 \sin^2{2 \beta}
  - \left( \frac{\lambda_\Delta^2}{8} + \frac{\lambda_S^2}{2} \right)
  v^2 \sin^2{2\beta}\, ,
\end{eqnarray}
respectively.

\subsection{Neutralino and Chargino Mass Matrices}

The fermionic components of the triplet and singlet superfields
mix with the MSSM neutralinos and charginos, 
and influence loop corrections to the mass of the Higgs boson.

In the basis of $\psi^0 = (\widetilde{B}, \widetilde{W}^0, \widetilde{h}_d^0,
\widetilde{h}_u^0, \widetilde{S}^0, \widetilde{\Delta}^0)$, the neutralino
mass matrix is given by
\begin{align}
M_{\widetilde{N}}
=
\left(
 \begin{array}{cccccc}
   M_1 & 0 & -\frac{g_Y}{2}v_d & \frac{g_Y}{2}v_u & 0 & 0\\
   0 & M_2 & \frac{g_2}{2}v_d & -\frac{g_2}{2}v_u & 0 & 0\\
   -\frac{g_Y}{2}v_d & \frac{g_2}{2}v_d & 0 & -\mu_{\rm eff} & -\frac{\lambda_S^{}}{\sqrt{2}}v_u & \frac{\lambda_\Delta^{}}{2\sqrt{2}}v_u\\
   \frac{g_Y}{2}v_u & -\frac{g_2}{2}v_u & -\mu_{\rm eff} & 0 & -\frac{\lambda_S^{}}{\sqrt{2}}v_d & \frac{\lambda_\Delta^{}}{2\sqrt{2}}v_d\\
   0 & 0 & -\frac{\lambda_s}{\sqrt2}v_u & -\frac{\lambda_s}{\sqrt2}v_d & \mu_s & 0\\
   0 & 0 & \frac{\lambda_\Delta^{}}{2\sqrt2}v_u & \frac{\lambda_\Delta^{}}{2\sqrt2}v_d & 0 &\mu_\Delta^{}   
 \end{array}
\right),
\end{align}
where the $M_1$ and $M_2$ denote the bino and wino masses, respectively.

In the basis of $\psi^+ = (\widetilde{W}^+, \widetilde{h}_u^+,
\widetilde{\Delta}^+)$ and $\psi^- = (\widetilde{W}^-, \widetilde{h}_d^-,
\widetilde{\Delta}^-)$, 
the chargino mass terms is given by
\begin{eqnarray}
  {\cal L} = - \frac{1}{2} (\psi^-)^T  M_{\widetilde{C}} \psi^+ 
  - \frac{1}{2} (\psi^+)^T  M_{\widetilde{C}}^T \psi^-\, ,
\end{eqnarray}
with
\begin{align}
M_{\widetilde{C}}
=
\left(
\begin{array}{ccc}
 M_2 & \frac{g}{\sqrt2} v_u  & 0\\
 \frac{g}{\sqrt2} v_d & \mu_{\rm eff} & \frac{\lambda_\Delta^{}}{2} v_u \\
 0 & -\frac{\lambda_\Delta^{}}{2} v_d & \mu_\Delta^{}
\end{array}
\right)\, .
\end{align}
This matrix is diagonalized by a bi-unitary transformation
\begin{align}
{\rm diag}(m_{\chi_i^+}) = U^\ast M_{\tilde{C}} V^\dagger\, ,
\end{align}
where the unitary matrices $U$ and $V$ rotate $\psi^-$ and
$\psi^+$ their corresponding mass eigenstates as
\begin{align}
\chi_i^- = U_{ij} \psi_j^-, \hspace{1em} 
\chi_i^+ = V_{ij} \psi_j^+.
\end{align}

\subsection{One Loop Corrections to the SM-like Higgs Boson Mass}

Here we discuss radiative corrections to the mass of the SM-like Higgs
boson at one loop level.  We follow the formalism described in
Refs.\cite{PBMZ,Degrassi:2009yq}, which takes the $\overline{\rm DR}$
scheme.  One loop corrected mass squared matrix for the CP-even Higgs
bosons in the gauge basis $S_i$ is given by
\begin{eqnarray}
  \left({\cal M}^2_S\right)^{\rm 1~loop}_{ij} 
  = \left( {\cal M}^2_S \right)^{\rm Tree} 
  + \frac{T_i}{v_i} \delta_{ij} - \Pi_{s_i s_j}(p^2)\, ,
\end{eqnarray}
where $T_i$ represent the finite part of the one loop tadpole
diagrams, and $\Pi_{s_i s_j}(p^2)$ the finite parts of the one loop
self-energy diagrams for external momentum $p$.  The form of
expressions for contributions to the scalar self-energies and tadpoles
are similar to those of the MSSM and NMSSM.  In our computation, we
include all contributions from MSSM particles to $\Pi_{s_1 s_1}$,
$\Pi_{s_2 s_2}$, $\Pi_{s_1 s_2}$, $T_1$ and $T_2$, and then add extra
contributions from extra Higgs, neutralino and chargino to
$\Pi_{s_2 s_2}$ and $T_2$.

The contributions to the scalar self energies from the Higgs bosons
loop diagrams are given by
\begin{align}
16\pi^2\Pi_{s_i s_j}^H(p^2) 
&= \sum_{k}^{4} 2 \lambda_{s_i s_j h_k h_k} A(m_{h_k}) 
+\sum_{k,\ell}^{4} 2\lambda_{s_i h_k h_\ell}\lambda_{s_j h_k h_\ell}B_0(m_{h_k},m_{h_\ell}) \nonumber\\
&+ \sum_{k}^{4} 2 \lambda_{s_i s_j a_k a_k} A(m_{a_k}) 
+\sum_{k,\ell}^{4} 2\lambda_{s_i a_k a_\ell}\lambda_{s_j a_k a_\ell}B_0(m_{a_k},m_{a_\ell}) \nonumber\\
&+ \sum_{k}^{4} 2 \lambda_{s_i s_j h^+_k h^-_k} A(m_{h_k^\pm}) 
+\sum_{k,\ell}^{4} \lambda_{s_i h_k^+ h_\ell^-}\lambda_{s_j h_k^+ h_\ell^-}B_0(m_{h_k^\pm},m_{h_\ell^\pm}).
\end{align}
The contributions to the scalar self-energies from neutralino and chargino loop
diagrams are given by
\begin{eqnarray}
16\pi^2\Pi^\chi_{s_i s_j}(p^2)
&=& 4 \sum_{k,\ell=1}^6 
{\rm Re}(\lambda_{s_i \chi^0_k \chi^0_\ell}\lambda_{s_j \chi^0_k \chi^0_\ell}^{\ast})
 \left[ (p^2 -m_{\chi^0_k}^2 -m_{\chi^0_\ell}^2 -2m_{\chi_k^0}m_{\chi_\ell^0} )
B_0( m_{\chi_k^0}, m_{\chi_\ell^0}) \right.  \nonumber \\
&& \hspace{5cm} \left. -A(m_{\chi^0_k}) -A(m_{\chi^0_\ell}) \right]\nonumber\\
&+& 2 \sum_{k,\ell=1}^3 {\rm Re}(\lambda_{s_i \chi^+_k \chi^-_\ell}\lambda_{s_j \chi^+_k \chi^-_\ell}^{\ast})
 \left[ (p^2 -m_{\chi^\pm_k}^2 -m_{\chi^\pm_\ell}^2 -2m_{\chi_k^\pm}m_{\chi_\ell^\pm} )B_0( m_{\chi_k^\pm}, m_{\chi_\ell^\pm}) \right. \nonumber \\
&& \hspace{5cm} \left. -A(m_{\chi^\pm_k}) -A(m_{\chi^\pm_\ell}) \right]\, 
\end{eqnarray}


The contributions to the tadpoles from Higgs boson loop diagrams are given by
\begin{align}
16\pi^2 T^\phi_i 
&= 
\sum_{\phi=h,a,h^\pm} \sum_{k=1}^{n_\phi} \lambda_{s_i \phi_k \phi_k} A(m_{\phi_k})\, ,
\end{align}
where $n_h = n_a = n_{h^\pm}^{} = 4$.
The contributions to the tadpoles from neutralino or chargino loop diagrams
are given by
\begin{align}
16\pi^2 T^\chi_i 
&= 
-4 \sum_{k=1}^{6} \lambda_{s_i \chi_k \chi_k} m_{\chi_k} A(m_{\chi_k}) 
-4 \sum_{k=1}^{3} \lambda_{s_i \chi^+_k \chi^-_k} m_{\chi_k^\pm} A(m_{\chi_k^\pm})\, .
\end{align}
Here, $A$ and $B_0$ are the Passarino-Veltman functions \cite{Passarino:1978jh}.
The tadpole and self-energy diagrams from SM fermions, gauge bosons
and fermions are similar to those of the MSSM, and we refer the
reader to \cite{PBMZ,Degrassi:2009yq}.

Definitions of the couplings $\lambda$ are given below.
Although we compute loop diagrams that contribute to mass shift in the top-left
$2 \times 2$ sub-matrix,
we list all CP-even Higgs couplings for completeness.

\subsubsection{Higgs self-couplings}

The trilinear self-couplings of the neutral Higgs bosons are given by
\begin{flalign}
&\lambda_{s_1 s_1 s_1} = \lambda_{s_1 p_1 p_1} = \frac{1}{8}(g_2^2 +g_Y^2)v_1,\hspace{1em}
\lambda_{s_2 s_2 s_2} = \lambda_{s_2 p_2 p_2} = \frac{1}{8}(g_2^2 +g_Y^2)v_2,\nonumber \\
&\lambda_{s_1 p_2 p_2} = 3\lambda_{s_1 s_2 s_2} = -\frac{1}{8}(g_2^2 +g_Y^2 -4\lambda_S^2 -\lambda_\Delta^2)v_1,\nonumber \\
&\lambda_{s_2 p_1 p_1} = 3\lambda_{s_1 s_1 s_2} = -\frac{1}{8}(g_2^2 +g_Y^2 -4\lambda_S^2 -\lambda_\Delta^2)v_2,\nonumber \\
&\lambda_{s_1 s_1 s_3} = \lambda_{s_2 s_2 s_3} = \frac{\lambda_S^{} }{3\sqrt2}\mu_{\rm eff},\hspace{1em}
\lambda_{s_1 s_1 s_4} = \lambda_{s_2 s_2 s_4} = \frac{\lambda_\Delta^{} }{6\sqrt2}\mu_{\rm eff},\nonumber \\
&\lambda_{s_1 p_3 p_3} = 3\lambda_{s_1 s_3 s_3} = \frac{\lambda_S^2}{2}v_1,\hspace{1em}
\lambda_{s_2 p_3 p_3} = 3\lambda_{s_2 s_3 s_3} = \frac{\lambda_S^2}{2}v_2,\nonumber \\
&\lambda_{s_1 p_3 p_4} = 3\lambda_{s_1 s_3 s_4} = \frac{\lambda_S^{} \lambda_\Delta^{}}{4}v_1,\hspace{1em}
\lambda_{s_2 p_3 p_4} = 3\lambda_{s_2 s_3 s_4} = \frac{\lambda_S^{} \lambda_\Delta^{}}{4}v_2,\nonumber \\
&\lambda_{s_1 p_4 p_4} = 3\lambda_{s_1 s_4 s_4} = \frac{\lambda_\Delta^2}{8}v_1,\hspace{1em}
\lambda_{s_2 p_4 p_4} = 3\lambda_{s_2 s_4 s_4} = \frac{\lambda_\Delta^2}{8}v_2,\nonumber \\
&\lambda_{s_3 p_1 p_1} = \lambda_{s_3 p_2 p_2} = \frac{\lambda_S^{}}{\sqrt2}\mu_{\rm eff},\hspace{1em}
\lambda_{s_4 p_1 p_1} = \lambda_{s_4 p_2 p_2} = \frac{\lambda_\Delta^{}}{2\sqrt2}\mu_{\rm eff},\nonumber \\
&\lambda_{s_1 s_2 s_3} = -\frac{\lambda_S^{}}{6\sqrt2}(A_S^{} +\mu_S^{}),\hspace{1em}
\lambda_{s_3 p_1 p_2} = \frac{\lambda_S^{}}{2\sqrt2}(A_S^{} +\mu_S^{}),\hspace{1em}
\lambda_{s_1 p_2 p_3} = \lambda_{s_2 p_1 p_3} = \frac{\lambda_S^{}}{2\sqrt2}(A_S^{} -\mu_S^{}),\nonumber \\
&\lambda_{s_1 s_2 s_4} = -\frac{\lambda_\Delta^{}}{12\sqrt2}(A_\Delta^{} +\mu_\Delta^{}),\hspace{1em}
\lambda_{s_4 p_1 p_2} = \frac{\lambda_\Delta^{}}{4\sqrt2}(A_\Delta^{} +\mu_\Delta^{}),\hspace{1em}
\lambda_{s_1 p_2 p_4} = \lambda_{s_2 p_1 p_4} = \frac{\lambda_\Delta^{}}{4\sqrt2}(A_\Delta^{} -\mu_\Delta^{})\, .
\end{flalign}
The quartic self-couplings of the neutral Higgs bosons are given by
\begin{flalign}
&\lambda_{s_1 s_1 s_1 s_1} = \lambda_{s_2 s_2 s_2 s_2} = \frac{1}{32}(g_2^2 +g_Y^2),\hspace{1em}
\lambda_{s_1 s_1 p_1 p_1} = \lambda_{s_2 s_2 p_2 p_2} = \frac{1}{16}(g_2^2 +g_Y^2),\nonumber \\
&\lambda_{s_1 s_1 s_2 s_2} = -\frac{1}{96}(g_2^2 +g_Y^2 -4\lambda_S^2 -\lambda_\Delta^2),\hspace{1em}
\lambda_{s_2 s_2 p_1 p_1} = \lambda_{s_1 s_1 p_2 p_2} = -\frac{1}{16}(g_2^2 +g_Y^2 -4\lambda_S^2 -\lambda_\Delta^2),\nonumber \\
&\lambda_{s_1 s_1 s_3 s_3} = \lambda_{s_2 s_2 s_3 s_3} = \frac{\lambda_S^2}{24},\hspace{1em}
\lambda_{s_1 s_1 p_3 p_3} = \lambda_{s_2 s_2 p_3 p_3} = \lambda_{s_3 s_3 p_1 p_1} = \lambda_{s_3 s_3 p_2 p_2} = \frac{\lambda_S^2}{4},\nonumber \\
&\lambda_{s_1 s_1 s_3 s_4} = \lambda_{s_2 s_2 s_3 s_4} = \frac{\lambda_S^{} \lambda_\Delta^{}}{48},\hspace{1em}
\lambda_{s_1 s_1 p_3 p_4} = \lambda_{s_2 s_2 p_3 p_4} = \frac{\lambda_S^{} \lambda_\Delta^{}}{4},\nonumber \\
&\lambda_{s_1 s_1 s_4 s_4} = \lambda_{s_2 s_2 s_4 s_4} = \frac{\lambda_\Delta^2}{96},\hspace{1em}
\lambda_{s_1 s_1 p_4 p_4} = \lambda_{s_2 s_2 p_4 p_4} = \lambda_{s_4 s_4 p_1 p_1} = \lambda_{s_4 s_4 p_2 p_2} = \frac{\lambda_\Delta^2}{16}\, .\nonumber \\
\end{flalign}
The trilinear couplings between the neutral and charged Higgs bosons are
written by
\begin{flalign}
&\lambda_{s_1^{} w_1^+ w_1^-} = \frac{1}{4}(g_2^2 +g_Y^2)v_d, \hspace{1em}
\lambda_{s_1^{} w_2^+ w_2^-} = \frac{1}{4}(g_2^2 -g_Y^2 +2\lambda_\Delta^2)v_d, \nonumber \\
&\lambda_{s_1^{} w_3^+ w_3^-} = \frac{g_2^2}{2}v_d, \hspace{1em}
\lambda_{s_1^{} w_4^+ w_4^-} = -\frac{1}{2}(g_2^2 -\lambda_\Delta^2)v_d, \nonumber \\
&\lambda_{s_2^{} w_1^+ w_1^-} = \frac{1}{4}(g_2^2 -g_Y^2 +2\lambda_\Delta^2)v_u, \hspace{1em}
\lambda_{s_2^{} w_2^+ w_2^-} = \frac{1}{4}(g_2^2 +g_Y^2)v_u, \nonumber \\
&\lambda_{s_2^{} w_3^+ w_3^-} = -\frac{1}{2}(g_2^2 -\lambda_\Delta^2)v_u, \hspace{1em}
\lambda_{s_2^{} w_4^+ w_4^-} = \frac{g_2^2}{2}v_u, \nonumber \\
&\lambda_{s_3^{} w_1^+ w_1^-} = \lambda_{s_3^{} w_2^+ w_2^-} = 
\sqrt{2} \lambda_S^{} \mu_{\rm eff} , \nonumber \\
&\lambda_{s_4^{} w_1^+ w_1^-} = \lambda_{s_4^{} w_2^+ w_2^-} = - \frac{\lambda_\Delta^{}}{\sqrt{2}}\mu_{\rm eff} ,
\nonumber \\
&\lambda_{s_1^{} w_1^+ w_2^-} = \frac{1}{8}(2g_2^2 -4\lambda_S^2 +\lambda_\Delta^2)v_u, \hspace{1em}
\lambda_{s_1^{} w_1^+ w_3^-} = \frac{1}{2}\lambda_\Delta^{} \mu_{\rm eff},\nonumber \\
&\lambda_{s_1^{} w_1^+ w_4^-} = \frac{\lambda_\Delta^{}}{2}\mu_{\rm eff}, \nonumber \\
&\lambda_{s_1^{} w_2^+ w_3^-} = \frac{\lambda_\Delta^{}}{2}\mu_{\Delta}^{}, \hspace{1em}
\lambda_{s_1^{} w_2^+ w_4^-} = \frac{\lambda_\Delta^{}}{2}A_{\Delta}^{}, \nonumber \\
&\lambda_{s_2^{} w_1^+ w_2^-} = \frac{1}{8}(2g_2^2 -4\lambda_S^2 +\lambda_\Delta^2)v_d, \hspace{1em}
\lambda_{s_2^{} w_1^+ w_3^-} = -\frac{\lambda_\Delta^{}}{2}A_{\Delta}^{}, \hspace{1em}
\lambda_{s_2^{} w_1^+ w_4^-} = -\frac{\lambda_\Delta^{}}{2}\mu_{\Delta}^{}, \nonumber \\
&\lambda_{s_2^{} w_2^+ w_3^-} = -\frac{\lambda_\Delta^{}}{2}\mu_{\rm eff}, \hspace{1em}
\lambda_{s_2^{} w_2^+ w_4^-} = -\frac{1}{2}\lambda_\Delta^{} \mu_{\rm eff}, \nonumber \\
&\lambda_{s_3^{} w_1^+ w_2^-} = \frac{\lambda_S^{}}{\sqrt{2}}(A_S^{} +\mu_S^{}), \hspace{1em}
\lambda_{s_3^{} w_1^+ w_3^-} = \lambda_{s_3^{} w_1^+ w_4^-} = \frac{\lambda_S^{} \lambda_\Delta^{}}{2\sqrt{2}}v_d^, \nonumber \\
&\lambda_{s_3^{} w_2^+ w_3^-} = \lambda_{s_3^{} w_2^+ w_4^-} = -\frac{\lambda_S^{} \lambda_\Delta^{}}{2\sqrt{2}}v_u^, \nonumber \\
&\lambda_{s_4^{} w_1^+ w_2^-} = -\frac{\lambda_\Delta^{}}{2\sqrt{2}}(A_\Delta^{} +\mu_\Delta^{}), \hspace{1em}
\lambda_{s_4^{} w_1^+ w_3^-} = -\lambda_{s_4^{} w_1^+ w_4^-} = -\frac{1}{4\sqrt{2}}(2g_2^2 -\lambda_\Delta^2)v_d, \nonumber \\
&\lambda_{s_4^{} w_2^+ w_3^-} = -\lambda_{s_4^{} w_2^+ w_4^-} = -\frac{1}{4\sqrt{2}}(2g_2^2 -\lambda_\Delta^2)v_u\, .\hspace{1em}
\end{flalign}
The quartic couplings between the neutral and charged Higgs bosons are given by
\begin{flalign}
&\lambda_{s_1^{} s_1^{} w_1^+ w_1^-} = \frac{1}{8}(g_2^2 +g_Y^2), \hspace{1em}
\lambda_{s_1^{} s_1^{} w_2^+ w_2^-} = \frac{1}{8}(g_2^2 -g_Y^2 +2\lambda_\Delta^2), \nonumber \\
&\lambda_{s_1^{} s_1^{} w_3^+ w_3^-} = \frac{g_2^2}{4}, \hspace{1em}
\lambda_{s_1^{} s_1^{} w_4^+ w_4^-} = -\frac{1}{4}(g_2^2 -\lambda_\Delta^2), \nonumber \\
&\lambda_{s_1^{} s_2^{} w_1^+ w_2^-} = \frac{1}{8}(2g_2^2 -4\lambda_S^2 +\lambda_\Delta^2), \nonumber \\
&\lambda_{s_1^{} s_3^{} w_1^+ w_3^-} = \lambda_{s_1^{} s_3^{} w_1^+ w_4^-} = \frac{\lambda_S^{} \lambda_\Delta^{}}{2\sqrt{2}}, \hspace{1em}
\lambda_{s_1^{} s_4^{} w_1^+ w_3^-} = -\lambda_{s_1^{} s_4^{} w_1^+ w_4^-} = -\frac{1}{4\sqrt{2}}(2g_2^2 -\lambda_\Delta^2), \nonumber \\
&\lambda_{s_2^{} s_2^{} w_1^+ w_1^-} = \frac{1}{8}(g_2^2 -g_Y^2 +2\lambda_\Delta^2), \hspace{1em}
\lambda_{s_2^{} s_2^{} w_2^+ w_2^-} = \frac{1}{8}(g_2^2 +g_Y^2), \nonumber \\
&\lambda_{s_2^{} s_2^{} w_3^+ w_3^-} = -\frac{1}{4}(g_2^2 -\lambda_\Delta^2), \hspace{1em}
\lambda_{s_2^{} s_2^{} w_4^+ w_4^-} = \frac{g_2^2}{4}, \nonumber \\
&\lambda_{s_2^{} s_3^{} w_2^+ w_3^-} = \lambda_{s_2^{} s_3^{} w_2^+ w_4^-} = -\frac{\lambda_S^{} \lambda_\Delta^{}}{2\sqrt{2}}, \hspace{1em}
\lambda_{s_2^{} s_4^{} w_2^+ w_3^-} = -\lambda_{s_2^{} s_3^{} w_2^+ w_4^-} = -\frac{1}{4\sqrt{2}}(2g_2^2 -\lambda_\Delta^2), \nonumber \\
&\lambda_{s_3^{} s_3^{} w_1^+ w_1^-} = \lambda_{s_3^{} s_3^{} w_2^+ w_2^-} = \frac{\lambda_S^2}{2}, \hspace{1em}
\lambda_{s_3^{} s_4^{} w_1^+ w_1^-} = \lambda_{s_3^{} s_4^{} w_2^+ w_2^-} = -\frac{\lambda_S^{} \lambda_\Delta^{}}{2}, \nonumber \\
&\lambda_{s_4^{} s_4^{} w_1^+ w_1^-} = \lambda_{s_4^{} s_4^{} w_2^+ w_2^-} = \frac{\lambda_\Delta^2}{8}, \hspace{1em}
\lambda_{s_4^{} s_4^{} w_3^+ w_3^-} = -\lambda_{s_4^{} s_4^{} w_3^+ w_4^-} = \lambda_{s_4^{} s_4^{} w_4^+ w_4^-} = \frac{g_2^2}{2}.
\end{flalign}

\subsubsection{Higgs couplings with neutralinos}

The couplings between CP-even Higgs bosons and neutralinos
are given by
\begin{align}
\cal{L}
&\supset -\sum_{i,k,\ell} \lambda_{s_i,\psi_k^0,\psi_\ell^0} S_i \psi_k^0 \psi_\ell^0
+{\rm h.c.},
\label{eq:Higgs-neutralino}
\end{align}
in terms of two component spinor notation.
The Higgs couplings with neutralinos are given by
\begin{align}
&\lambda_{s_1 \psi_1 \psi_3} = -\frac{g_Y^{}}{4}, \hspace{1em}
\lambda_{s_1 \psi_2 \psi_3} = +\frac{g_2^{}}{4}, \hspace{1em} \nonumber \\
&\lambda_{s_1 \psi_4 \psi_5} = +\frac{\lambda_S^{}}{2\sqrt{2}}, \hspace{1em}
\lambda_{s_1 \psi_4 \psi_6} = +\frac{\lambda_\Delta^{}}{4\sqrt{2}}, \hspace{1em} \nonumber \\
&\lambda_{s_2 \psi_1 \psi_4} = +\frac{g_Y^{}}{4}, \hspace{1em}
\lambda_{s_2 \psi_2 \psi_4} = -\frac{g_2^{}}{4}, \hspace{1em} \nonumber \\
&\lambda_{s_2 \psi_3 \psi_5} = -\frac{\lambda_S^{}}{2\sqrt{2}}, \hspace{1em}
\lambda_{s_2 \psi_3 \psi_6} = +\frac{\lambda_\Delta^{}}{4\sqrt{2}}, \hspace{1em} \nonumber \\
&\lambda_{s_3 \psi_3 \psi_4} = -\frac{\lambda_S^{}}{2\sqrt{2}}, \hspace{1em}
\lambda_{s_4 \psi_3 \psi_4} = +\frac{\lambda_\Delta^{}}{4\sqrt{2}}.
\end{align}
In the neutralino mass eigenstates $\chi_i^0$, 
their couplings to the CP-even Higgs boson $s_i$ are given by
\begin{align}
\lambda_{s_i \chi_k^0 \chi_\ell^0} = N_{ka}^\ast N_{lb}^\ast \lambda_{s_i \psi_a^0 \psi_b^0}, 
\end{align}
where $N$ is the diagonalization matrix for neutralino mass matrix.

\subsubsection{Higgs couplings with charginos}

The couplings between the CP-even Higgs bosons and charginos are given by
\begin{align}
\cal{L} 
&\supset -\sum_{i, k, \ell} \lambda_{s_i \psi_k^+ \psi_\ell^-} S_i \psi_k^+ \psi_\ell^-
+{\rm h.c.},
\end{align}
where $\psi_i^+ = (\widetilde{W}^+,\widetilde{h}_u^+,\widetilde{\Delta}^+)$ and
$\psi_i^- = (\widetilde{W}^-,\widetilde{h}_d^-,\widetilde{\Delta}^-)$. 
The Higgs couplings with charginos are given by
\begin{align}
&\lambda_{s_1 \psi_1^+ \psi_2^-} = \frac{g_2^{}}{\sqrt{2}}, \hspace{1em}
\lambda_{s_1 \psi_2^+ \psi_3^-} = \frac{\lambda_\Delta^{}}{2}, \hspace{1em} \nonumber \\
&\lambda_{s_2 \psi_3^+ \psi_2^-} = \frac{\lambda_\Delta^{}}{2}, \hspace{1em} 
\lambda_{s_2 \psi_2^+ \psi_1^-} = \frac{g_2^{}}{\sqrt{2}}, \hspace{1em} \nonumber \\
&\lambda_{s_3 \psi_2^+ \psi_2^-} = \frac{\lambda_S^{}}{\sqrt{2}}, \hspace{1em}
\lambda_{s_4 \psi_1^+ \psi_3^-} = \frac{g_2^{}}{4}, \hspace{1em} \nonumber \\
&\lambda_{s_4 \psi_2^+ \psi_2^-} = \frac{\lambda_\Delta^{}}{2\sqrt{2}}, \hspace{1em} 
\lambda_{s_4 \psi_3^+ \psi_1^-} = -\frac{g_2^{}}{4}\, . 
\end{align}

\end{document}